\documentclass{ws-ijmpb}
\usepackage[sort&compress,comma,super]{natbib}
\setcitestyle{super,open={},close={}}

\def\sgn{\mathop{\textrm{sgn}}} 
 
\newcommand{\beq}{
\begin{equation}
	} 
	\newcommand{\eeq}{
\end{equation}
} 
\newcommand{\be}{
\begin{eqnarray}
	} 
	\newcommand{\ee}{
\end{eqnarray}
}

\newcommand{\dg}{{\dagger}} 
\newcommand{\pdg}{{\vphantom\dagger}} 
\newcommand{\px}{{p_x}} 
\newcommand{\bQ}{{\bf Q}} 
\newcommand{\bp}{{\bf p}} 
\newcommand{\py}{{p_y}} 
\newcommand{\bea}{
\begin{eqnarray}
	} 
	\newcommand{\eea}{
\end{eqnarray}
} 
\newcommand{\bG}{{\bf G}} 
\newcommand{\bK}{{\bf K}}
\newcommand{\bk}{{\bf k}}

\begin{document}

\markboth{M. Killi, S. Wu, and A. Paramekanti}{Graphene: Kinks, Superlattices, Landau Levels, and Magnetotransport}

%%%%%%%%%%%%%%%%%%%%% Publisher's Area please ignore %%%%%%%%%%%%%%%
%
\catchline{}{}{}{}{}
%
%%%%%%%%%%%%%%%%%%%%%%%%%%%%%%%%%%%%%%%%%%%%%%%%%%%%%%%%%%%%%%%%%%%%

\title{Graphene: Superlattices, Topological Kinks, Landau Levels and Tunable Magnetotransport
%\footnote{For the title, try not to 
%use more than 3 lines. Typeset the title in 10 pt 
%Times roman, uppercase and boldface.}
}

\author{\footnotesize Matthew Killi}
%\footnote{Typeset names in
%10~pt Times roman, uppercase. Use the footnote to indicate 
%the present or permanent address of the author.}}

\address{Department of Physics, University of Toronto, Toronto, Ontario, 
Canada M5S 1A7\\
mkilli@physics.utoronto.ca}

\author{Si Wu}

\address{Department of Physics and Astronomy, University of Waterloo, Ontario, 
Canada N2L 3G1\\
Department of Physics, University of Toronto, Toronto, Ontario, 
Canada M5S 1A7\\
si.wu@uwaterloo.ca}

\author{Arun Paramekanti}

\address{Department of Physics, University of Toronto, Toronto, Ontario, 
Canada M5S 1A7 \\
Canadian Institute for Advanced Research, Toronto, Ontario, Canada M5G 1Z8 \\
arunp@physics.utoronto.ca}

\maketitle

\begin{history}
\received{(Day Month Year)}
\revised{(Day Month Year)}
\end{history}

\begin{abstract} We review recent work on superlattices in monolayer and bilayer graphene. We highlight the role of the
quasiparticle chirality in generating new Dirac fermion modes with tunable anisotropic velocities in one dimensional (1D)
superlattices in both monolayer and bilayer graphene. We discuss the structure of the Landau levels and magnetotransport in such
superlattices over a wide range of perpendicular (orbital) magnetic fields. In monolayer graphene, we show that an orbital
magnetic field can reverse the anisotropy of the transport imposed by the superlattice potential, suggesting possible
switching-type device applications. We also consider topological modes localized at a kink in an electric field applied
perpendicular to bilayer graphene, and show how interactions convert these modes into a two-band Luttinger liquid with tunable
Luttinger parameters. The band structures of electric field superlattices in bilayer graphene (with or without a magnetic field)
are shown to arise naturally from a coupled array of such topological modes. We briefly review some bandstructure results for 2D
superlattices. We conclude with a discussion of recent tunneling and transport experiments and point out open issues.
\end{abstract}


\section{Introduction}

Graphene is a two-dimensional carbon crystal that exhibits novel physics and transport properties due to its excitations
resembling chiral relativistic massless Dirac fermions at low energy.\cite{Castro-Neto:2009,DasSarma:RMP2011,Abergel:AdvPhy2012}
Its bilayer cousin, Bernal-stacked bilayer graphene (BLG) has also garnered much interest due to the possible novel broken
symmetry states it could potentially exhibit in the presence of interactions that destabilize the quadratic band touching point
present in a minimal tight-binding
model.\cite{Lemonik:2010,Zhang:2010,Nandkishore:2010a,Nandkishore:2010b,Vafek:2010,Nandkishore:2012,MacDonald:2012,
Miransky:2010,Miransky:2011} Both graphene
and bilayer graphene are also widely regarded as viable materials for developing new types of device applications due to the
chiral nature of their low energy excitations. This is largely due to the possibility of opening tunable band gaps - by
engineering a relative potential difference on each sublattice \cite{Zhou:2007,Giovannetti:2007,Kindermann:2012a} 
or by strain engineering \cite{Low:2010,Low:2011} in monolayer
graphene or by applying an electric field perpendicular to the layers in
bilayer graphene.\cite{McCann:2006a,Castro:2007,Zhang:2009}

In this review, we focus on new physics that develops in the presence of slow spatial potential variations in both monolayer
and bilayer graphene. While a spatially varying chemical potential could result in p-n junctions in either case,\cite{Lau:NJP2009, Lau:NanoLett2010} a perpendicular
electric field that changes direction as a function of position, specific to bilayer graphene, leads to midgap domain wall
states (or `kink' states) that have topological character.\cite{Martin:2008,Xavier:2010,Jung:2011a} For an isolated ``nanowire'',
electron interactions drive this 1D system into a tunable two-band Luttinger liquid.\cite{Killi:2010} These tunable nanowires have also been shown
to act as controllable ``electronic highways''.\cite{Qiao:2011} In both monolayer and bilayer graphene, profiles with {\it
periodic} potential variations form superlattices that can lead to dramatic modifications in the bulk band structure.  Such superlattice potentials have been shown to induce
new Dirac modes at zero and finite energy with tunable velocities and transport
properties.\cite{Tan:2011,Barbier:2008,Brey:2009,Park:2008,Park:2008a,Park:2008b,Pletikosic:2009,Barbier:2010b,Arovas:2010,Rusponi:2010,Barbier:2010a,Killi:2011a,Burset:2011a,Yankowitz:2012}

We review the physics of such superlattices as well as the effect of a perpendicular orbital magnetic field for three
experimentally accessible field regimes: weak, moderate, and strong fields.\cite{Wu:2012a} A weak orbital magnetic field
essentially acts as a `probe' of the Dirac modes, while a strong magnetic field overwhelms the effect of the superlattice
potential leading to quantum Hall physics indistinguishable from that of pure graphene. At moderate magnetic fields,
however, we find dispersing Landau levels and interesting field tuning of transport properties. In addition, we discuss the
effect of a magnetic field on the so-called topological `kink' states that form at the interface that separates two region with
opposite interlayer bias.\cite{Zarenia:2011a,Zarenia:2011,Wu:2012a,Huang:2012} Throughout this article we discuss recent
developments, ongoing experimental efforts, and open issues.

\section{Superlattices (SLs) in monolayer graphene (MLG)}

\subsection{Bandstructure of 1D superlattices}
For pristine MLG, ignoring spin,
the low energy Hamiltonian is given by a $2\times 2$ matrix at each valley, $ H_0 \! = \! v_f (s p_x \sigma_x - p_y \sigma_y)$, where pseudospin $\sigma_z=\pm 1$ labels the two trigonal sublattices, while the two (decoupled) valleys at $\pm \bK=\pm 4\pi \hat{x}/{\sqrt{3} a}$ are labelled by $s \! = \! \pm 1$. Here, $v_f \! = \! 3ta/2$ is the isotropic Fermi velocity, with $a \! = \! 1.42$ \, \AA and $t \! = \! 3$eV being the nearest neighbor carbon-carbon distance and transfer integral respectively, $\bp$ is the momentum measured from $\bK$. (We set $\hbar \! =\! 1$ for convenience.) The quasiparticles
in the vicinity of each valley then behave as massless linearly dispersing Dirac fermions,
with an energy dispersion $v_f |{\bf p}|$.

We focus here on the effect of a smooth SL potential, for which the period of the SL is significantly larger than the interatomic distance. This means we can safely ignore the intervalley scattering of electrons which involves large momentum transfer. We therefore use the above low energy Hamiltonian and focus on the electronic properties of SLs near a single valley. To this end, a 1D SL potential can be modelled as $H_{SL}=U(y)I$, where $U(y)=U(y+\lambda)$ with $\lambda$ being the SL period, and $I$ is the identity matrix in the pseudospin (sublattice) space. To gain a qualitative
understanding of the nontrivial phenomena arising from such SLs, 
such as the anisotropic Fermi velocity renormalization or SL induced band gaps,
%zone boundary (MBZ),
it is not necessary to assume any specific form for $U(y)$.
%SL potential $U(y)$ can be taken to be a step function, $U(y)=U_0\sgn(y)$ for $|y|<\lambda/2$, or to %have a cosine form, $U(y)=U_0\cos(y)$. As we will see later, 
%a s$V(y)$, anisotropic renormalization of Fermi velocities, gap opening at the mini Brillouin zone %boundary (MBZ), etc.

The problem of finding the energy spectrum of $H=H_0+H_{SL}$ has been extensively studied. It 
was noticed by Park {\it et. al.}\cite{Park:2008} that the 
chiral nature of Dirac fermions in graphene leads to the anisotropic renormalization of Fermi 
velocity near Dirac point. Surprisingly, the Fermi velocity is 
not renormalized in the SL direction, but is suppressed perpendicular to the modulation direction,
a counterintuitive effect that is deeply rooted in the chiral
nature of the Dirac fermions in MLG.

\subsubsection{Weak SL Potential}

For a weak SL potential, the energy spectrum near Dirac point and Fermi velocity renormalization can be well understood from perturbation theory. By expanding the Hamiltonian $H$ in the chiral basis, $|\bp s\rangle=\frac{1}{\sqrt{2}}(1,s{\rm e}^{-i\theta_{\bk}})^T$, where $\cos\theta_{\bp}=p_x/|\bp|$ and $s=\pm$ denotes electron and hole states, the kinetic energy part $H_0$ can be brought into diagonal form, while the matrix elements between $|\bp s\rangle$ and $|\bp+n\bG, s^{\prime}\rangle$, with $\bG=(0,2\pi/\lambda)$ as the reciprocal lattice vector, is given by $\frac{U(n\bG)}{2}(1+ss^{\prime}{\rm e}^{i\theta_{\bp,\bp+n\bG}})$, where $\theta_{\bp,\bp+n\bG}=\theta_{\bp}-\theta_{\bp+n\bG}$. Therefore, for states with momenta parallel to the SL direction, $\bp=(0,p_y)$, we can show that the full Hamiltonian matrix will consists of two decoupled blocks, 
\begin{equation}
	\label{Matrix}
	\left(
	\begin{array}{cccccccccc}
		\ddots & & & & & & & & & \\
		& \varepsilon_{e}({\bf p}-{\bf G}) & U({\bf G}) & U(2{\bf G}) & & & & & & \\
		& U^*({\bf G}) & \varepsilon_{e}({\bf p}) & U({\bf G}) & & & & & & \\
		& U^*(2{\bf G}) & U^*({\bf G}) & \varepsilon_{h}({\bf p}+{\bf G}) & & & & & & \\
		& & & & \ddots & & & & & \\
		& & & & & & \varepsilon_{e}({\bf p}+{\bf G}) & U({\bf G}) & U(2{\bf G}) & \\
		& & & & & & U^*({\bf G}) & \varepsilon_{h}({\bf p}) & U({\bf G}) & \\
		& & & & & & U^*(2{\bf G}) & U^*({\bf G}) & \varepsilon_{h}({\bf p}-{\bf G}) & \\
		& & & & & & & & & \ddots
	\end{array}
	\right),
\end{equation}
where $\varepsilon_{e,h}(\bk)=\pm v_F|\bk|$ are the electron (hole) energies. Applying the second order perturbation theory, the energy correction at momentum $\bk$ is given by
\begin{equation}
	\label{2ndEnergy}
	\Delta E^{(2)}=\sum_{n\neq 0}\frac{\left|U(n{\bf G})\right|^2}
	{\varepsilon({\bf p})-\varepsilon({\bf p}-n{\bf G})},
\end{equation} 
where the sum is carried out in the same block and $\varepsilon({\bf p})$ is understood as the corresponding electron or hole energies. This term is zero because of the linearity of the spectrum, {\it i.e.}, $\varepsilon({\bf p})-\varepsilon({\bf p}-n{\bf G})=\varepsilon({\bf p}+n{\bf G})-\varepsilon({\bf p})$. This means, along the direction of the superlattice, the Fermi velocity is not renormalized. Therefore, the absence of Fermi velocity renormalization in the SL direction is a consequence of the chiral nature of Dirac electrons and the linearity of the spectrum.

At the MBZ boundary, $\bp=(0,\pm\pi/\lambda)$, the two blocks become exactly identical, which means the energy will be doubly degenerate and energy spectrum is gapless at this point. This result is exact and independent of perturbation theory, and this band touching point will always be present.

\begin{figure}[t]
	\centering
	\includegraphics[width=.46\textwidth]{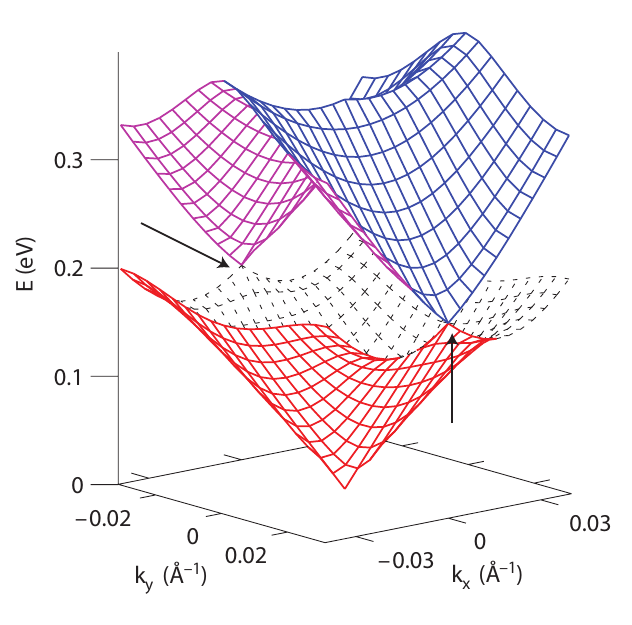}
	\includegraphics[width=.46\textwidth]{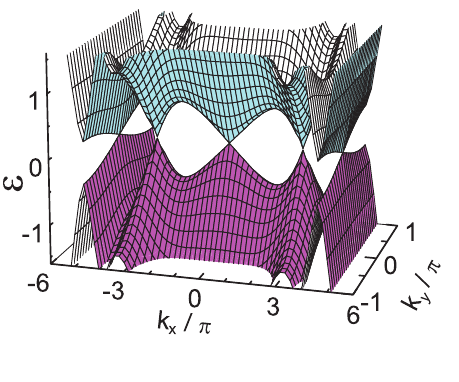}
	\caption{Left: Energy spectrum for a single anisotropic Dirac cone.
	[Reprinted by permission from Macmillan Publishers Ltd:
	Nature Physics {\bf 4}, 213, (2008).]	Notice the robust band crossing and nontrivial minigap opening at the MBZ boundary.
	Right:
	Energy spectrum showing
	five anisotropic Dirac cones. [Reprinted with permission from 
	Ref.\cite{Barbier:2010a}. Copyright (2010) American Physical Society]. 
	}
	\label{DiracFig}
\end{figure}

Once the momentum $\bp$ is no longer parallel to the reciprocal lattice vector $\bG$, this nice decoupling will break down and Fermi velocity in the corresponding direction, $v_{\hat{p}}={\bf v}(\bp)\cdot\hat{p}$, will inevitably become renormalized. Within second order perturbation approximation, the renormalization with respect to pristine MLG is\cite{Park:2008}
\beq
	\frac{v_{\hat{p}}-v_f}{v_f}=-\sum_{n\neq 0}\frac{2|U(n\bG)|^2}{n^2v_f^2|\bG|^2}\sin^2\theta_{\bp,\bG},
	\label{anire}
\eeq
where $\theta_{\bp,\bG}$ is the angle between $\bp$ and $\bG$. Since the right hand side of Eq. (\ref{anire}) is always negative, except for $\bp \parallel \bG$, the Fermi velocity is decreased from pristine MLG value. In contrast, for artificial electrons with linear dispersion but no chirality in a weak 1D SL, the second order perturbation result for the Fermi velocity is given by 
\beq
	\frac{v_{\hat{p}}-v_f}{v_f}=-\sum_{n\neq 0}\frac{2|U(n\bG)|^2}{n^2v_f^2|\bG|^2},
\eeq
which is isotropically decreased and is independent of the direction of $\bp$. Therefore, the anisotropic renormalization of the Fermi velocity near Dirac cone is truly a signature of chiral low energy excitations in MLG.  The left Figure \ref{DiracFig} shows the dispersion for a weak 1D SL potential, and corroborates the above results.  

\begin{figure}[t]
	\centering
	\includegraphics[width=.8\textwidth]{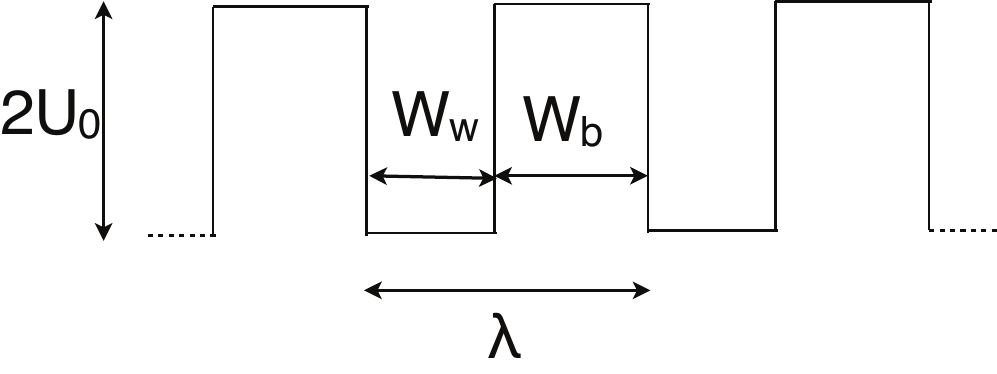}
	\caption{Step SL potential, where $V_0$ is the SL potential strength, $\lambda$ is the SL period, and $W_w(W_b)$ is the width of potential well (barrier).}
	\label{SL}
\end{figure}
For a square barrier SL (Fig.\ref{SL}), the energy spectrum can be found exactly.\cite{Barbier:2010a,Arovas:2010} By making use of Bloch theorem and matching boundary condition, it can be shown that the energy spectrum can be obtained from the following transcendental equation,
\beq
	\cos p_x=\cos(\lambda_wl_w)\cos(\lambda_bl_b)-Q\sin(\lambda_wl_w)\sin(\lambda_bl_b).
	\label{spectrum}
\eeq
Here, we have used the following notation:
\bea
	\varepsilon_w=\varepsilon+ul_b,\ \ \ \varepsilon_b=\varepsilon-ul_w,\ \ \ 
	u=\frac{U_0\lambda}{v_f},\ \ \ l_{b,w}=\frac{W_{b,w}}{\lambda},\nonumber\\
	\lambda_w=\left(\varepsilon_w^2-p_x^2\right)^{1/2},\ \ \ 
	\lambda_b=\left(\varepsilon_b^2-p_x^2\right)^{1/2}\ \ \ {\rm and}\ \ \ 
	Q=\frac{\varepsilon_w\varepsilon_b-p_x^2}{\lambda_w\lambda_b}.
\eea
For a symmetric SL, $W_b=W_w$, or equivalently $l_b=l_w=\frac{1}{2}$, Eq. (\ref{spectrum}) reduces to
\beq
	\cos p_x=\cos\frac{\lambda_w}{2}\cos\frac{\lambda_b}{2}
	-Q\sin\frac{\lambda_w}{2}\sin\frac{\lambda_b}{2},
	\label{symm}
\eeq
where $\varepsilon_w=\varepsilon+u/2$ and $\varepsilon_b=\varepsilon-u/2$. For this symmetric case, the energy spectrum is particle-hole symmetric, which means Eq. (\ref{symm}) is invariant under the transformation $\varepsilon\rightarrow -\varepsilon$.

To obtain the behavior near $\bK$ point, we can expand Eq. (\ref{symm}) in small $\varepsilon$ and $p_x$. The result is 
\beq
	\varepsilon=\pm\left(4\sin^2\left(\frac{p_y}{2}\right)+\frac{p_x^2\sin^2(u/4)}
	{(u/4)^2}\right)^{1/2}.
\eeq
From this, we can see that the low energy spectrum is indeed described by an anisotropic Dirac cone, with $v_y=v_f$ and 
\beq
	v_x=v_f\frac{\sin(u/4)}{u/4}.
\eeq

\subsubsection{Strong SL potential}

Since the Fermi velocity perpendicular to the SL direction can be significantly renormalized and even brought to zero for a broad region in momentum space, the energy spectrum becomes dispersionless in this direction and the electrons can be collimated in the SL direction.\cite{Park:2008a} Moreover, extra Dirac points can be generated in the energy spectrum for an even stronger 1D SL, which are shown in the right panel of Fig.~\ref{DiracFig}.\cite{Barbier:2010a,Arovas:2010,Park:2008b,Brey:2009} 

To determine the condition for the emergence of extra Dirac points and also their locations, we consider a symmetric SL and assume $p_y=0$ and $\varepsilon=0$ in Eq. (\ref{symm}). Then Eq. (\ref{symm}) reduces to 
\beq
	1=\cos^2\left(\frac{\lambda_w}{2}\right)+\frac{u^2/4+p_x^2}{u^2/4-p_x^2}\sin^2\left(\frac{\lambda_w}{2}\right),
\eeq
which can be solved by either $u^2/4+p_x^2=u^2/4-p_x^2$ or $\sin^2(\lambda_w/2)=0$. The former condition gives $p_x=0$, which is just the original Dirac point. The latter condition leads to $\lambda_w/2=j\pi$ with $j$ being a nonzero integer. Then, the position of new Dirac points are subsequently found at
\beq
	p_x=\pm\sqrt{\frac{u^2}{4}-4j^2\pi^2}.
\eeq
For asymmetric SL, the energy spectrum is no longer particle hole symmetric and extra Dirac points will appear with nonzero energies. 

At the induced Dirac points, the Fermi velocities behave differently from the original Dirac point.\cite{Barbier:2010a} To see this, we can expand Eq. (\ref{symm}) in $\varepsilon$ up to second order and obtain
\beq
	\varepsilon_{\pm}=\pm\left[\frac{4|a^2|^2[k_x^2\sin^2(a/2)+a^2\sin^2(k_y/2)]}
	{k_x^4a\sin a+a^2u^4/16-2k_x^2u^2\sin^2(a/2)}\right]^{1/2},
\eeq
with $a=[u^2/4-k_x^2]^{1/2}$. Then, at the $j$th extra Dirac point, the Fermi velocities along $x$ and $y$ directions are given by
\beq
	v_x=\frac{u^2/4-4j^2\pi^2}{u^2}v_f,\ \ \ v_y=\frac{16\pi^2j^2\cos(k_y/2)}{u^2}v_f.
\eeq
In contrast, at the original Dirac point, $v_x=4v_f\sin(u/4)/u$ and $v_y=v_f$.

%\subsection{Local density of states (at zero field)?}

\begin{figure}[t]
	\centering
	\includegraphics[width=.9\textwidth]{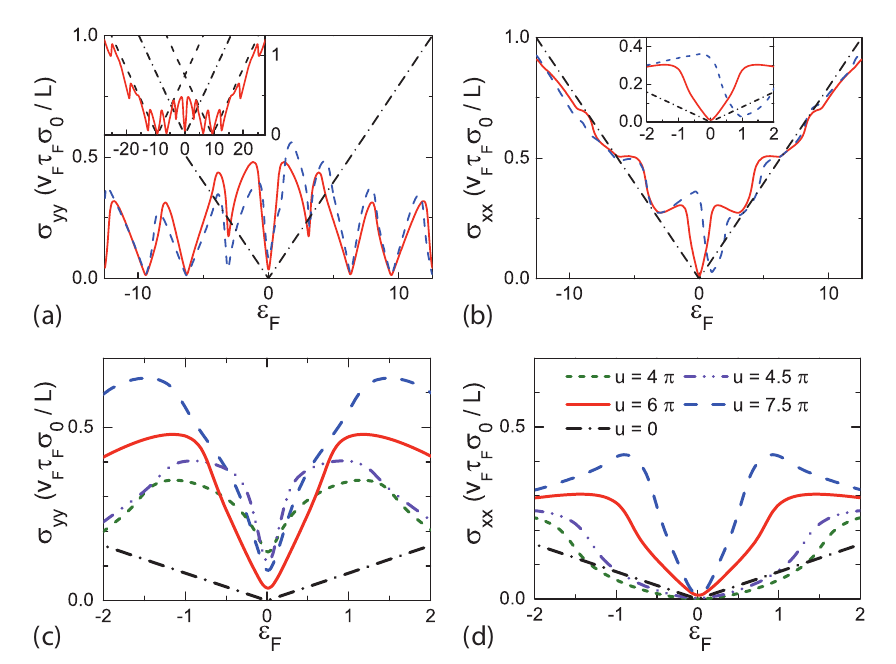}
	\caption{(a), (b) $\sigma_{yy}$ and $\sigma_{xx}$ as a function of Fermi energy for an SL with $u=6\pi$, for $l_b=0.5$ (red curve) and $l_b=0.4$ (dashed blue curve) respectively. 
	(c),(d) $\sigma_{yy}$ and $\sigma_{xx}$ as a function of Fermi energy with $l_b=0.5$ for different SL potential strength. 
	[Figures reprinted with permission from 
	Ref.\cite{Barbier:2010a}. Copyright (2010) American Physical Society.]}
	\label{Zero}
\end{figure}

\subsection{Zero field transport}

By assuming a constant relaxation time at the Fermi energy $\tau(E_F)=\tau_F$, the dc conductivity for MLG SL can be calculated by\cite{Barbier:2010a}
\beq
	\sigma_{ii}(E_F)=\frac{e^2\beta\tau_F}{A}\sum_{n,\bk}v_{ni}^2f_{n\bk}(1-f_{n\bk}),
\eeq
where $v_{ni}=\langle n\bk|v_i|n\bk\rangle$ is the average velocity in the $i$-th direction for $n$-th energy band, $f_{n\bk}=1/\{\exp[\beta(E_{n\bk}-E_F)]+1\}$ is the Fermi-Dirac distribution with $\beta=1/k_BT$.

The results for the various conductivities as a function of Fermi energy are shown in Fig. \ref{Zero}. Fig. \ref{Zero}(a) and \ref{Zero}(b) show $\sigma_{yy}$ and $\sigma_{xx}$, respectively, for an SL with $u=6\pi$ and $\beta=v_f/k_BT\lambda=20$. Here, red and dashed blue curve correspond to symmetric ($l_b=0.5$) and asymmetric ($l_b=0.4$) SL, respectively. From Fig. \ref{Zero}(a), notice that $\sigma_{yy}$ is oscillating when the Fermi energy is below the barrier but increases on the average almost linearly when the Fermi energy is above the barrier. On the other hand, $\sigma_{xx}$ always increases on the average with the Fermi energy. Both $\sigma_{yy}$ and $\sigma_{xx}$ show oscillating behavior and are symmetric (asymmetric) for symmetric (asymmetric) SLs. For $\sigma_{yy}$, however, there is a dip at the crossing energies of those mini bands. From Eq. (\ref{spectrum}), it can be shown that crossing energies occur at $\varepsilon=n\pi$ at $p_x=0$. 

Fig. \ref{Zero}(c) and \ref{Zero}(d) show $\sigma_{yy}$ and $\sigma_{xx}$ respectively for a symmetric SL with different SL potential strength. Notice that, at low energies, $\sigma_{xx}$ is smaller than its value in the absence of an SL. Upon increasing SL potential strength, $\sigma_{xx}$ also increases due to the appearance of extra Dirac points. Also, we can notice that $\sigma_{xx}$ is always smaller than $\sigma_{yy}$ at low energies, since $v_y>v_x$ near the Dirac point.

Now let us consider conductivities for symmetric SLs at zero temperature and charge neutrality ({\it i.e.},$T=0$ and $E_F=0$) with only one Dirac point in the spectrum.  As the SL potential is not very strong, the low energy Hamiltonian can be written as $H=v_xk_x\sigma_x+v_yk_y\sigma_y$, with anisotropic Fermi velocities. It can be shown that the conductivity along and perpendicular to the SL direction is given by
\beq
	\sigma_{yy}(E_F=0)=\frac{v_x}{v_y}\sigma_0=\sigma_0\frac{|\sin(u/4)|}{u/4},\ \ \ 
	\sigma_{xx}(E_F=0)=\frac{v_y}{v_x}\sigma_0=\sigma_0\frac{u/4}{|\sin(u/4)|},
	\label{single}
\eeq
where $\sigma_0$ is the universal conductivity of an isotropic Dirac cone. The value of $\sigma_0$ depends on the order of different limits taken, such as vanishing temperature and Fermi energy. However, the form of the result for an anisotropic Dirac cone does not depends how $\sigma_0$ is calculated.

When there are extra Dirac points in the spectrum, by assuming their independence and using Eq. (\ref{single}), the conductivities now are
\bea
	&&\sigma_{yy}(E_F=0)=\sigma_0\left(\frac{|\sin(u/4)|}{u/4}+2\sum_{j=1}^{j_{{\rm max}}}
	+\frac{(u/4)^2-j^2\pi^2}{j^2\pi^2}\right),\nonumber\\
	&&\sigma_{xx}(E_F=0)=\sigma_0\left(\frac{u/4}{|\sin(u/4)|}+2\sum_{j=1}^{j_{{\rm max}}}
	+\frac{j^2\pi^2}{(u/4)^2-j^2\pi^2}\right),
	\label{multi}
\eea
where $j_{{\rm max}}={\rm Int}[u/4\pi]$ counts pairs of extra Dirac points. From Eq. (\ref{multi}) we can see, every time a new pair of Dirac points is generated in the spectrum with $u=4n\pi$ and $n$ an integer, the conductivity parallel to the SL will shows a dip, while conductivity in the perpendicular direction will diverge.

\begin{figure}[t]
	\centering
	\includegraphics[width=.46\textwidth]{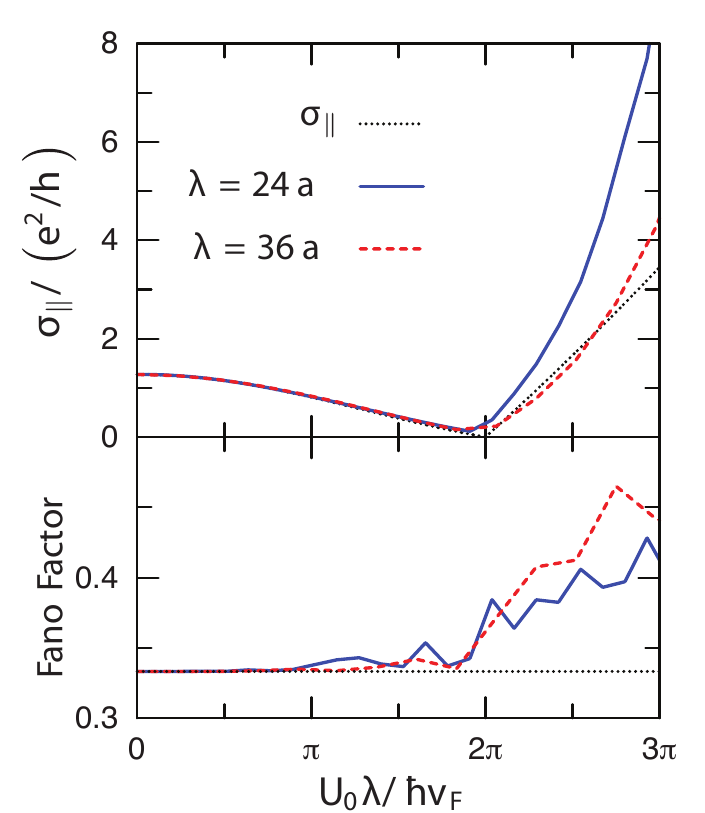}
	\includegraphics[width=.46\textwidth]{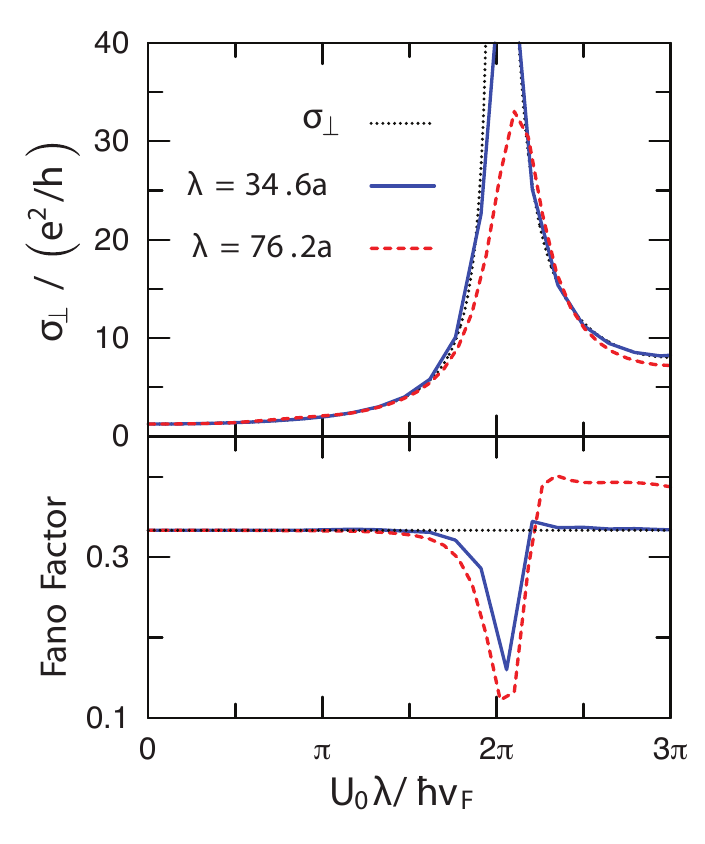}
	\caption{Conductivity parallel (left) and perpendicular (right) to the SL direction and the corresponding Fano factors. Here, $\lambda$ is the SL period and $a=1.42$ \AA is the lattice constant of graphene. The black dotted line corresponds to conductivity calculated from Eq. (\ref{multi}) (top panel) and Fano factor $F=1/3$ (bottom panel).
	[Reprinted with permission from 
	Ref.\cite{Burset:2011a}. Copyright (2011) American Physical Society.]
	}
	\label{Landauer}
\end{figure}

These results can be confirmed by numerical calculation, using Landauer-B\"{u}ttiker formalism.\cite{Burset:2011a} Fig. \ref{Landauer} shows the corresponding numerical results. In the left, the conductivity parallel to the SL direction (top panel) agrees with Eq. (\ref{single}) when $U_0$ is smaller than the critical value where extra Dirac points are generated. The corresponding Fano factor is 1/3 (bottom panel), which agrees with pseuodiffusive character of transport. Once $U_0$ exceeds the critical value, new Dirac points will emerge and provide new transmission channels in the SL direction. We can see that the conductivity qualitatively agrees with Eq. (\ref{multi}). Since Eq. (\ref{multi}) is based on the assumption that all the Dirac points are independent and have linear dispersion, which is valid in a very small energy region, it is not surprising to see the numerical results depends on both SL potential strength and SL period. Also, the Fano factor is larger than 1/3, indicating that the transport is no longer pseudodiffusive.

In contrast, conductivity perpendicular to the SL direction is shown in the right figure of Fig. \ref{Landauer}. Again, when SL potential strength is smaller than the critical value, the conductivity is well described by the simple picture of Eq. (\ref{single}). When $U_0$ approaches the critical value, the conductivity shows a peak. For even stronger SL potential, the numerical result agrees with Eq. (\ref{multi}) quite well, which may suggest that the approximations adopted for Eq. (\ref{multi}) are appropriate in the perpendicular direction. Also, the Fano factor is 1/3 for almost all SL potential strengths, except for those critical values where new Dirac points are generated.

\subsection{Landau levels}

In a uniform perpendicular magnetic field, the eigenenergy for pristine MLG in the absence of SL is $\varepsilon_n={\rm sgn}(n)\sqrt{|n|}\omega_c$, where $\omega_c \! =\! \sqrt{2}v_F/\ell_B$, with $\ell_B \! =\! 1/\sqrt{eB}$. For $s=+1$ (i.e., at valley ${\bK}$), the $n \! \neq \! 0$ eigenfunctions are given by \be &&\phi_{n,k,+} (x,y)=\frac{e^{ i k x}}{\sqrt{2 L}}\left( 
\begin{array}{c}
	\psi_{|n|,k}(y) \\
	- {\rm sgn}(n) \psi_{|n|-1,k}(y) 
\end{array}
\right), \label{eigen1} \ee where $L$ and $k$ are the system length and electron momentum deviation from $\bK$, both along the $x$-direction, while for $n=0$, \beq \phi_{0,k,+} (x,y)=\frac{e^{ikx}}{\sqrt{L}} \left( 
\begin{array}{c}
	\psi_{0,k}(y) \\
	0 
\end{array}
\right). \label{eigen2} \eeq Here, $\psi_{n,k}(y)$ is the n-th eigenstate of a (shifted) 1D harmonic oscillator, \beq \psi_{n,k}(y)=\frac{1}{\sqrt{2^{n}n!\sqrt{\pi}\ell_B}}{\rm exp} \left[-\frac{1}{2}\left(\frac{y-y_0}{\ell_B}\right)^2\right]
H_{n}\left(\frac{y-y_0}{\ell_B}\right), \eeq centered at $y_0 \! = \! k\ell_B^2$, and $H_n$ are Hermite polynomials. For $s\!=\!-1$ (i.e., at $- \bK$), the eigenfunctions are given by $\phi_{n,k,-}(x,y) \!=\! -i \sigma_y \phi_{n,k,+}(x,y)$. The full low energy LLs of MLG are thus $\phi_{n,k,\pm} (x,y) {\rm e}^{\pm i K_x x}$.

We now turn to the effect of a periodic 1D chemical potential modulation $V(y)$, with period $\lambda \! \gg \! a$, on these Landau levels at low energy. Recent work has shown these results to be generally
consistent with solving the Harper equation using the full tight-binding model.\cite{Katsnelson:2012}
 The set of eigenfunctions $\phi_{n,k,s} (x,y) {\rm e}^{ i s K_x x}$, with $s=\pm 1$, form a convenient basis to study the SL Hamiltonian in a magnetic field. (This basis choice is different from the one used by Park, {\it et al},\cite{Park:2009} and allows us to numerically access a wide range of magnetic fields. \cite{Schweitzer:2012}
 In the weak field regime, our results are consistent with Ref.~\cite{Park:2009}.) 
 Due to momentum conservation along the $x$-direction, the SL Hamiltonian is diagonal in $k$. Further, for $\lambda \gg a$, intervalley scattering is strongly suppressed. We will therefore assume that the two valleys stay completely decoupled. (We focus below on valley ${\bK}$ with $s\!=\! +1$; we expect identical physics around valley $-{\bK}$.) With this approximation, the only effect of the SL potential is, thus, to induce Landau level mixing.

To proceed, we need to choose a concrete form for the SL potential. For simplicity, we set $V(y) \! =\! \frac{U}{2}\cos\left(\frac{2\pi y}{\lambda}\right)$, although our results can be easily generalized to other (e.g., step-like) SL potentials by including multiple Fourier components. We can then expand the Hamiltonian in the above basis, retaining up to 3000 Landau levels, and diagonalize it to obtain the spectrum of the 1D SL in a magnetic field. 

In order to study the effect of the magnetic field on the 1D SL in graphene, with $\tilde{U} =U\lambda/2\pi v_f \sim {\cal O}(1)$, it is useful to consider three regimes for the magnetic field.
\begin{figure}[t] 
	\centering
	\includegraphics[width=.4
	\textwidth]{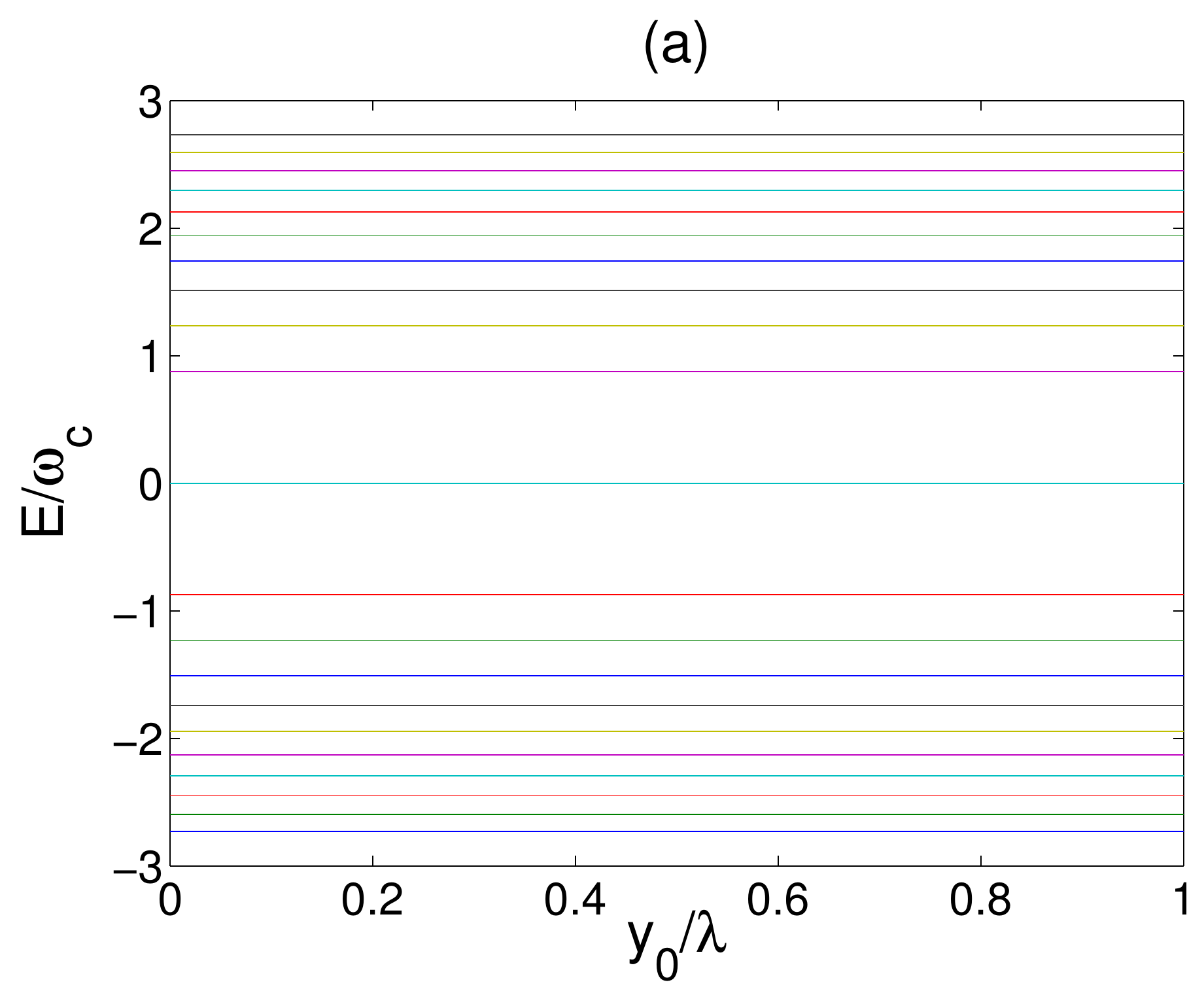} 
	\includegraphics[width=.4
	\textwidth]{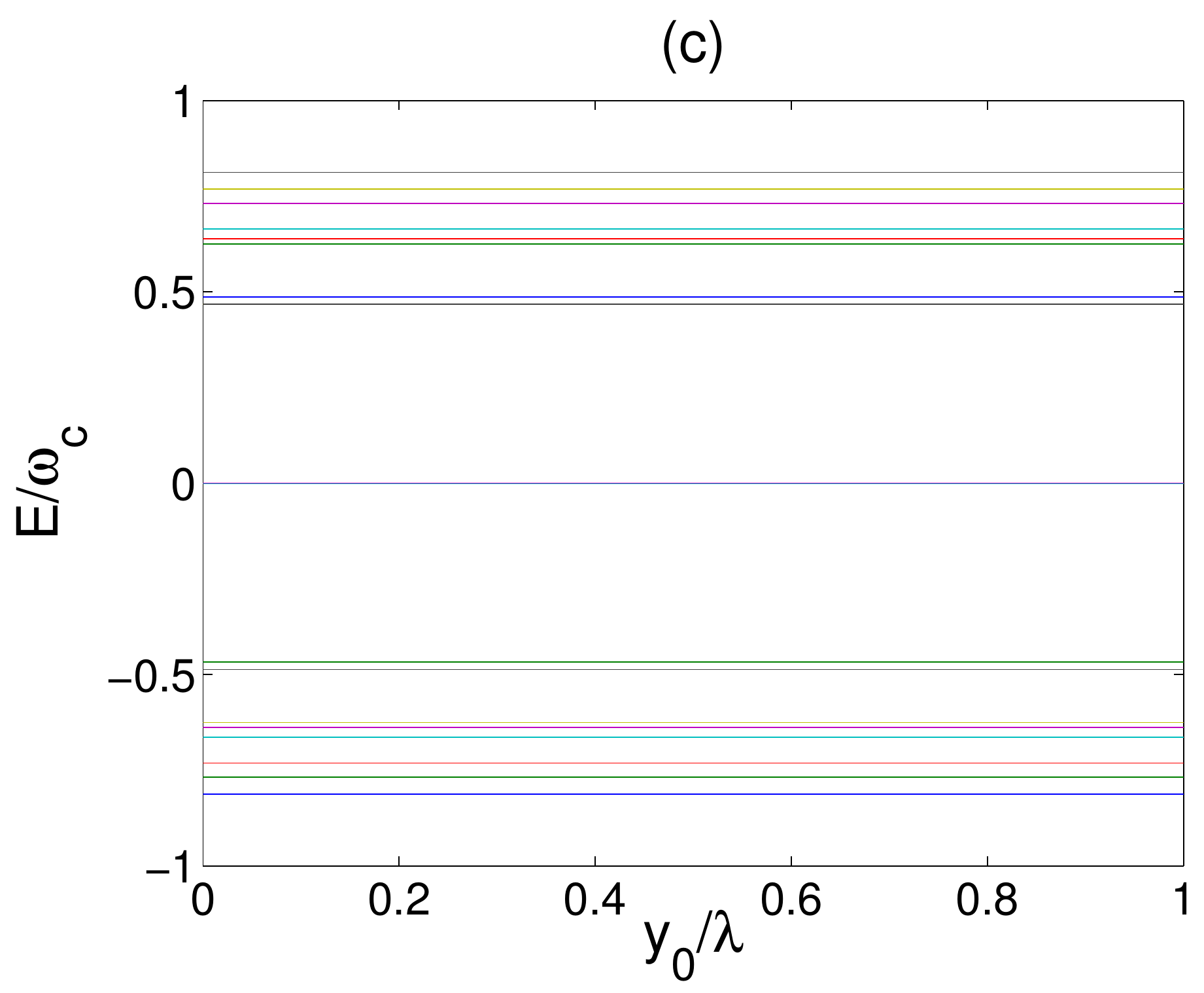} 
	\includegraphics[width=.4
	\textwidth]{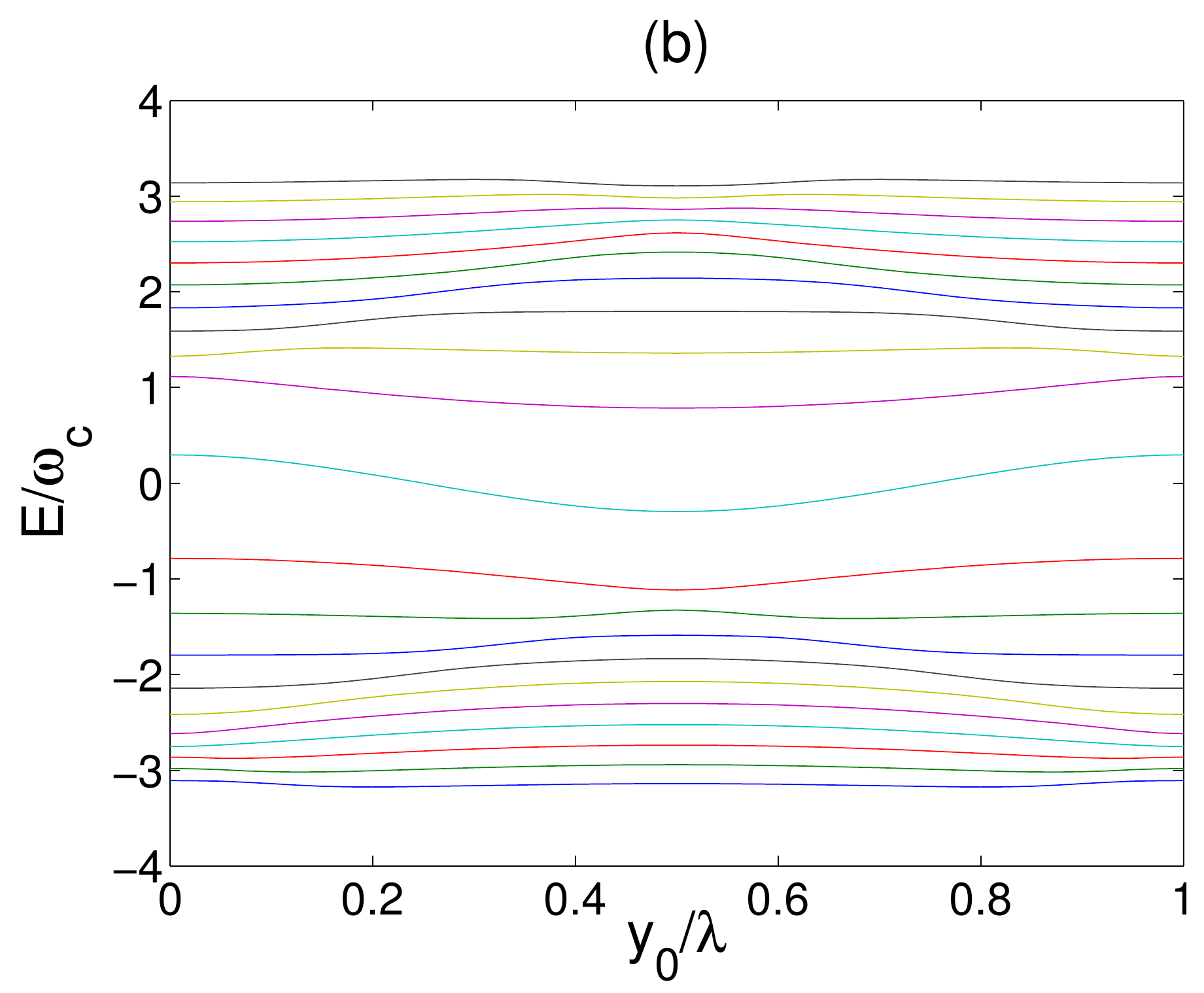} 
	\includegraphics[width=.4
	\textwidth]{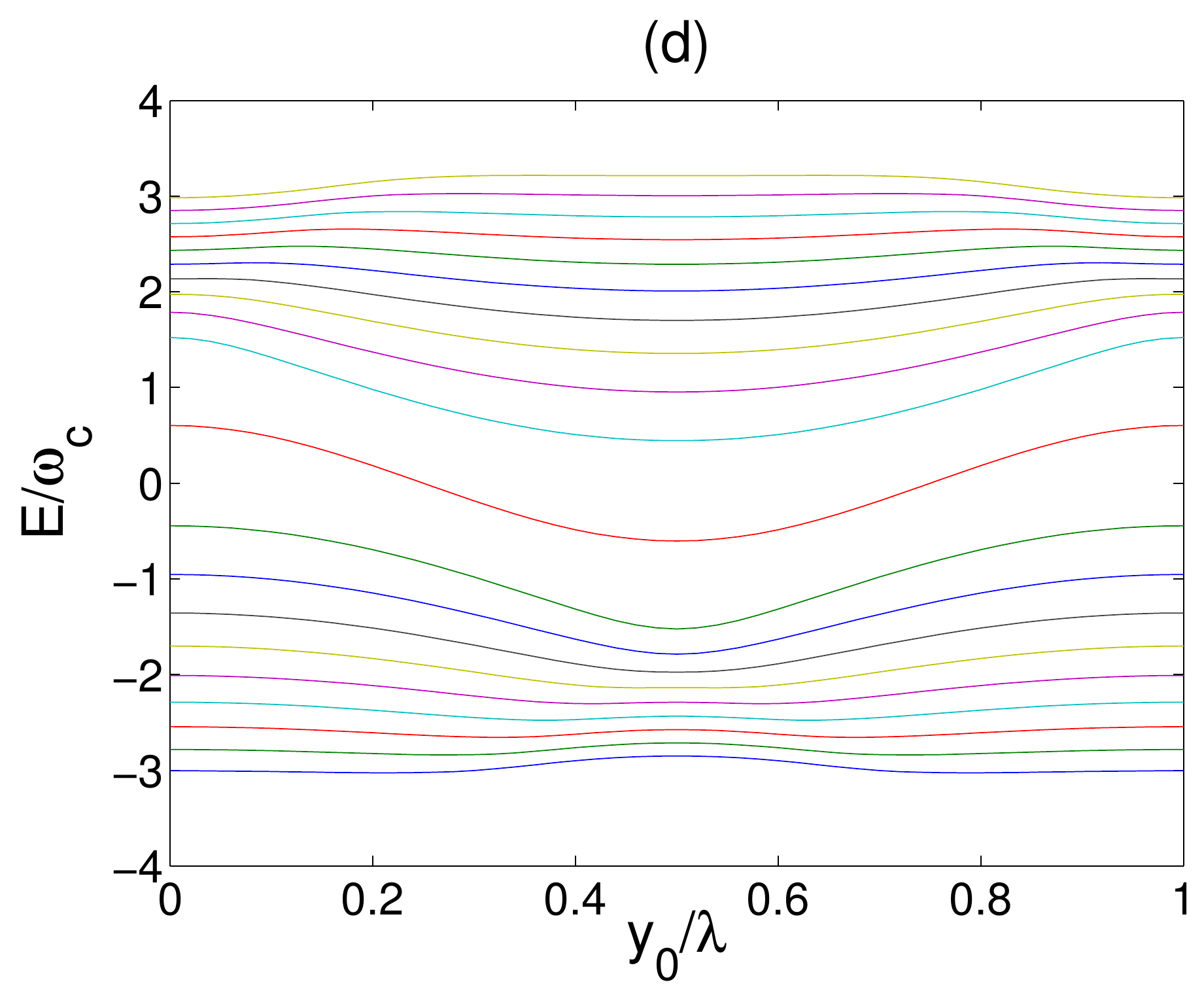} 
	\caption{(Color online) Landau levels of monolayer graphene SL for different (dimensionless) SL strengths $\tilde{U}$, and magnetic fields $B$. The spectrum is shown for weak field ($\ell_B=2\lambda$, top panels) and intermediate field ($\ell_B=0.2\lambda$, bottom panels). Left panels (a,b) correspond to $\tilde{U}=1$ which supports a single anisotropic zero energy massless Dirac fermion. Right panels (c,d) correspond to $\tilde{U}=3$ which supports three zero energy massless Dirac fermions with anisotropic velocities - the weak field zero energy LL thus has three times as many states for $\tilde{U}=3$ as it does for $\tilde{U}=1$, while the $n=\pm 1, \pm 2$ levels have degeneracy splitting in weak field due to the Dirac fermions having two different mean velocities. For $\ell_B \ll \lambda$ (not shown), the LLs closely resemble that of pristine graphene. See text for a detailed discussion of the Landau level structure.} 
	\label{GSL_LL} 
\end{figure}

(i) {\bf Weak field}: This regime corresponds to having $\hbar \omega_c \ll U$, where the Landau level spacing is much smaller than the SL amplitude, so that $2 \pi \ell_B/\lambda \gg 1$. In this regime, the magnetic field may be viewed as effectively `probing' the zero field SL excitations.

(ii) {\bf Intermediate field}: In this regime, $\hbar \omega_c \sim U$, which means $2\pi \ell_B/\lambda \sim 1$, so that the SL potential and the magnetic field have to be treated on equal footing.

(iii) {\bf Strong field}: Here, $\hbar \omega_c \gg U$ or, equivalently, $2\pi \ell_B/\lambda \ll 1$. In this regime, the SL potential only weakly perturbs the Landau levels of pristine graphene.

Fig.~\ref{GSL_LL} shows the spectrum of the graphene SL in different field regimes for SL strengths $U\!\!=\!\!2\pi v_f/\lambda$ (or $\tilde{U}=1$) and $6\pi v_f/\lambda$ (or $\tilde{U}=3$). This allows us to contrast the behaviour of the spectrum of the SL in a magnetic field without or with extra Dirac points being present at zero field, and to explore consequences for quantum Hall physics and transport.

\subsubsection{Weak field regime} \label{section:weak}

When the magnetic field is weak, $\ell_B=2\lambda$ (top panels in Fig. \ref{GSL_LL}), we find that the energy spectrum barely depends on the value of $k$, or equivalently, $y_0$. This is due to the fact that when magnetic length $\ell_B$ is larger than the SL period $\lambda$, the matrix elements of the Hamiltonian do not depend on the center of the LL wavefunctions, which yields flat bands. Equivalently, in this regime, the magnetic field may be viewed as effectively `probing' the structure of the zero field SL dispersion leading to Landau levels which depend on the nature of the Dirac spectrum at low energy.

For $\tilde{U}=1$, the low energy spectrum of the SL contains a single anisotropic Dirac point at zero energy. For an anisotropic Dirac cone described by an effective Hamiltonian $H=v_xk_x\sigma_x+v_yk_y\sigma_y$, the LLs are given by $\varepsilon_n=\sgn(n)\sqrt{2|n|v_xv_y}/\ell_B$. Since the SL renormalizes $v_x < v_f$, but leaves $v_y=v_f$, the Landau levels at weak field resemble those of pristine graphene, but with a renormalized lower effective velocity $\sqrt{v_x v_y} < v_f$.

For $\tilde{U}=3$, the low energy spectrum of the SL contains three anisotropic Dirac points at zero energy, so that the zero energy Landau level has three times the degeneracy of the case with $\tilde{U}=1$. Further, the Dirac cone centred at $\bK$ has a slightly different average velocity $\sqrt{v_x v_y}$ compared with the two cones which are symmetrically split off from $\bK$ along $\pm \hat{x}$. This degeneracy breaking results in the Landau levels at nonzero energy becoming weakly split, as is most clearly seen for the first two excited Landau levels (at positive or negative energy, i.e., with $n=\pm 1, \pm 2$). We have numerically determined $v_x$ and $v_y$ for each of the three Dirac points and found good agreement between the energy levels obtained on this basis of having Dirac fermions with two different average velocities, and that obtained directly numerically.

At higher energies, $E/\omega_c \gtrsim 2$ for $\tilde{U}=1$ or $E/\omega_c \gtrsim 1$ for $\tilde{U}=3$, the spectrum begins to deviate from this simple behavior expected for a linear Dirac spectrum. This deviation results from curvature in the dispersion, which appears upon going beyond the linearized approximation.

\subsubsection{Intermediate field regime} 
\label{section:inter}

At intermediate fields, for $\ell_B=0.2\lambda$, the spectrum at low energy is most simply understood as arising from the SL potential inducing a strong dispersion to the Landau levels. In simple terms, if we assume that the state labelled by momentum $k$, or equivalently position $y_0$, have an energy which is modulated by the SL potential, we expect a periodic modulation of this energy with period $\lambda$ and amplitude proportional to the SL amplitude $U$. The behaviour of the low energy Landau levels, $n=0,\pm 1, \pm 2$, as seen from the lower panels in Fig.\ref{GSL_LL}, is consistent with this scenario, with the modulation following the $\cos(2\pi y/\lambda)$ form of the SL potential and the modulation for $\tilde{U}=3$ being roughly thrice as strong as the modulation for $\tilde{U}=1$. We can also see that the low energy Landau levels when $\tilde{U}=3$ overlap with each other. This will have nontrivial effect on the dc conductivity, as shown in the following subsection. For higher energy Landau levels, the energy spectrum still has a periodic modulation but no longer retains the simple form of cosine function. This is due to the fact that as the energy gets higher, the distribution of Landau levels becomes more dense and the energy difference between two adjacent levels is now comparable to the matrix element of SL potential. Therefore, a simple first order perturbation correction is not enough to account for the dispersion and second order perturbation from adjacent levels must be taken in account, which causes the Landau level to lose its simple cosine form.

\subsubsection{High field regime}
\label{section:high}

For very strong magnetic field, the Landau level structure of pristine graphene is recovered. Here, only one zero energy level exists at the Dirac point, and other energy levels follow the square root relation. This is simply because in such a strong magnetic field, the SL is just a perturbation and can only give rise to a small modulation of the LLs following our argument at intermediate field. From a perturbative point of view, the energy corrections up to first order to the LL energies are given by \beq \Delta E^{(1)}=\int dy \phi_n^*(k,y)V(y)\phi_n(k,y), \eeq which gives a sinusoidal dependence on the center position of LL wavefunctions. Thus, even in a strong magnetic field, the energy spectrum is not dispersionless but has a spatial modulation following the SL; however the ratio of the amplitude of this modulation to the Landau level spacing, $U/\omega_c$, is extremely small in the high field regime. This dispersion, though small, can give rise to interesting magnetoresistance oscillation known as Weiss oscillation, on top of the usual Shubnikov-de Hass oscillation. \cite{Matulis} It was shown that, compared to two-dimensional electron gas with parabolic dispersion relation, Weiss oscillation in graphene SL is more pronounced and is more robust against temperature damping in small field region. This is a consequence of the different Fermi velocities of Dirac and normal electrons at same chemical potential. \cite{Matulis} 

\subsection{Magnetotransport}
\label{section:conductivity}

\begin{figure}
	[t] \centering 
	\includegraphics[width=.4
	\textwidth]{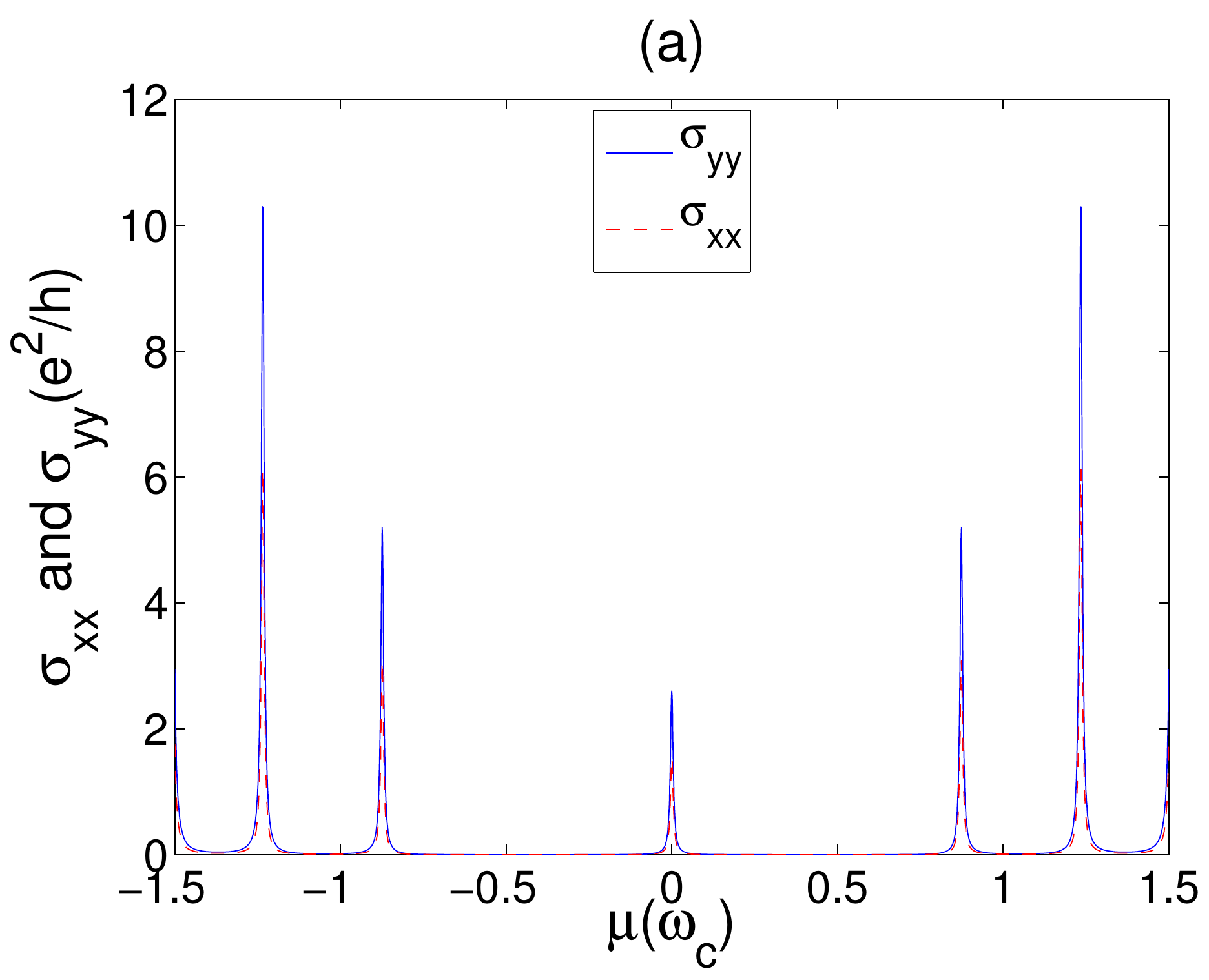} 
	\includegraphics[width=.4
	\textwidth]{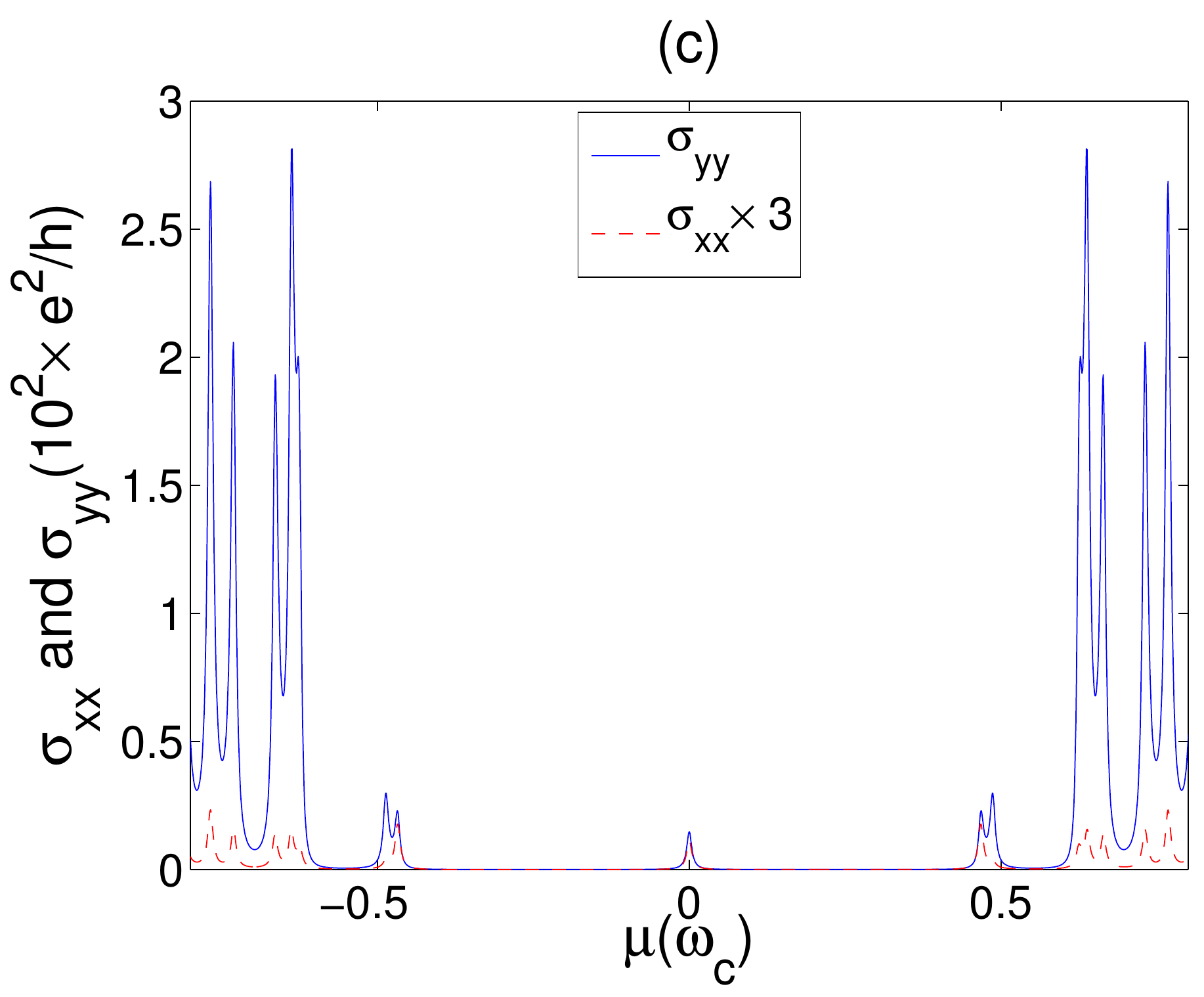} 
	\includegraphics[width=.4
	\textwidth]{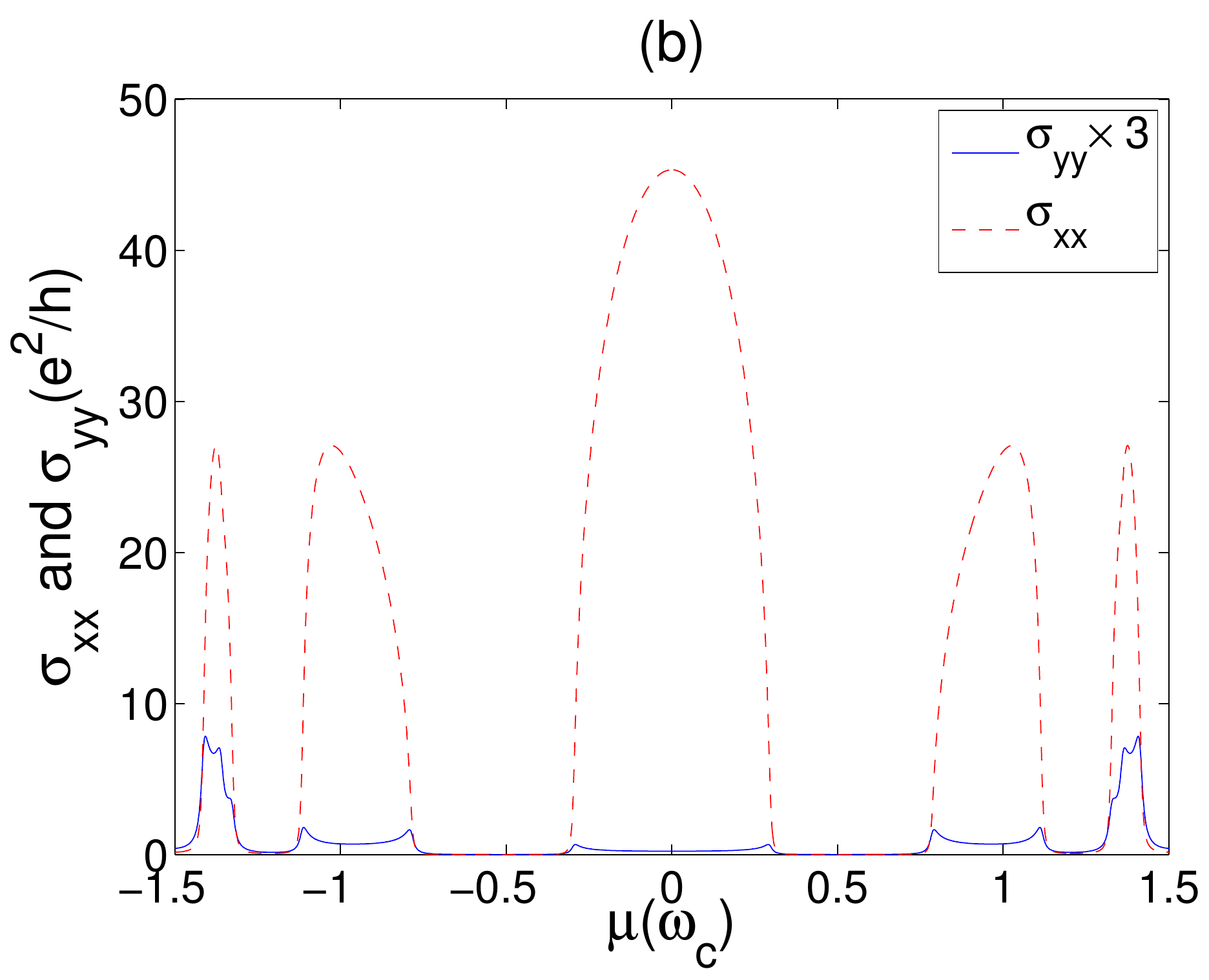} 
	\includegraphics[width=.4
	\textwidth]{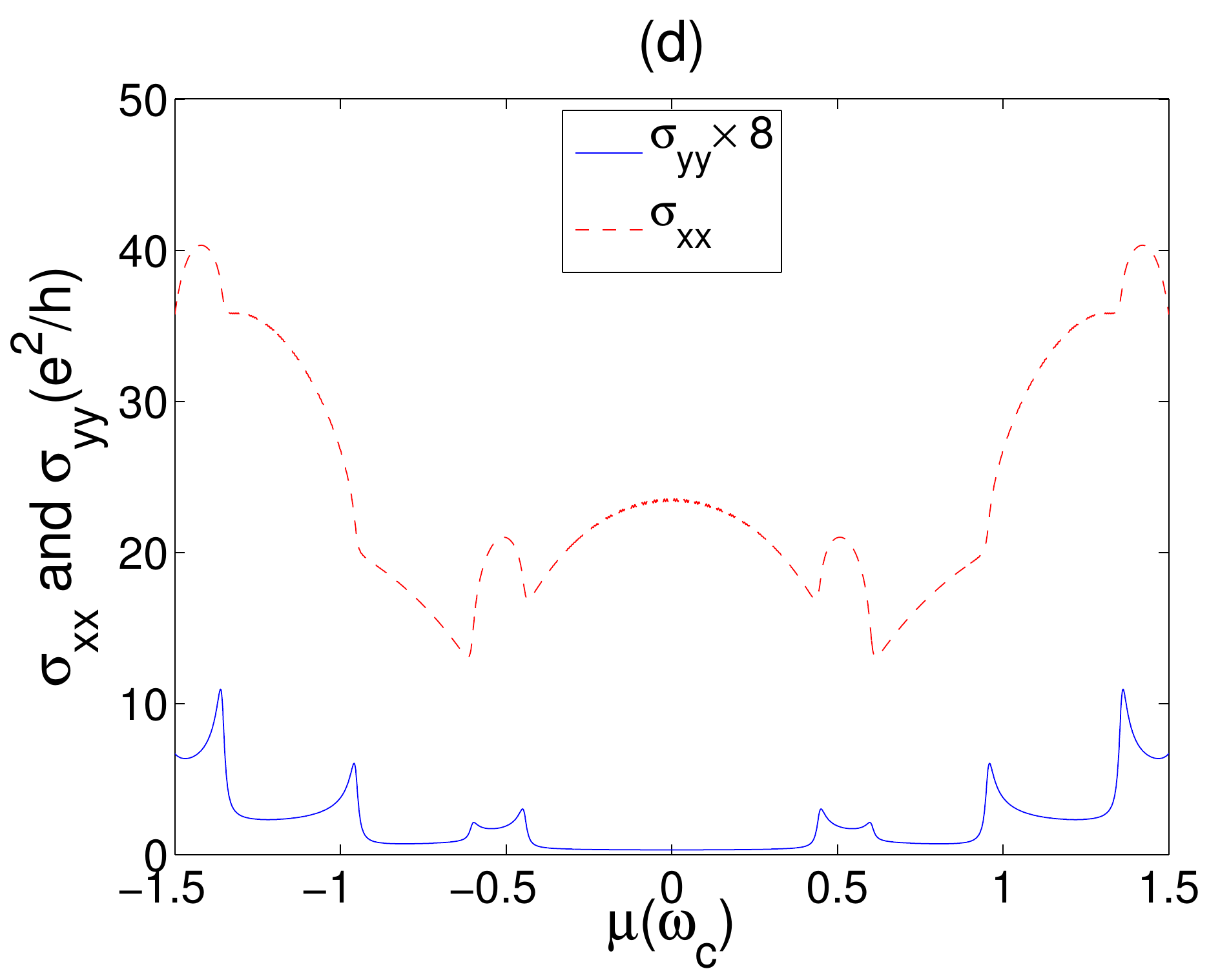} \caption{ (Color online) Diagonal dc conductivities of monolayer graphene SL for different (dimensionless) SL strengths $\tilde{U}$, and magnetic fields $B$. The conductivity is shown for weak field ($\ell_B=2\lambda$, top panels) and intermediate field ($\ell_B=0.2\lambda$, bottom panels). Left panels (a,b) correspond to $\tilde{U}=1$, and right panels (c,d) correspond to $\tilde{U}=3$. The conductivities show strong anisotropy when magnetic field strength is tuned - for weak field (a,c), $\sigma_{yy}$ is larger than $\sigma_{xx}$, which is a consequence of the Fermi velocity renormalization in the absence of magnetic field; for moderate field (b,d), the anisotropy is reversed, since $\hat{v}_x$ acquires intra-LL contributions, as explained in the text. For $\ell_B \ll \lambda$ (not shown), result for pristine graphene is recovered and the transport is isotropic in both directions.} \label{diagonal_SLG} 
\end{figure}

Once we have the eigenvalues and eigenfunctions for the superlattice in a perpendicular magnetic field, both ac and dc conductivities can be calculated directly by Kubo formula, 
\beq 
\sigma_{ij}(\omega)=\frac{e^2}{h}\frac{1}{\pi\lambda\ell^2}\int_0^{\lambda}dy_0\sum_{\alpha,\beta} 
\frac{f(E_{\alpha})-f(E_{\beta})}{E_{\alpha}-E_{\beta}}\frac{\langle\alpha k |v_i|\beta k\rangle\langle\beta k |v_j|\alpha k\rangle}
{E_{\alpha}-E_{\beta}-\omega-i\Gamma}. 
\label{Kubo}
\eeq  Here, we have set $\Gamma=10^{-3}\times 2\pi v_f/\lambda$ as the Landau level broadening, $E_{\alpha}(y_0)$ and $|\alpha k\rangle$ are the $\alpha$-th eigenvalue and the corresponding eigenstate of the system which can be expanded in the basis of $|nk\rangle$, where $\phi_n(k,y)=\langle y|nk\rangle$ is the LL wavefunctions for pristine graphene. $v_i$ is the velocity operator in $\hat{i}$-direction and $v_i=v_F\sigma_i$, where $\sigma_i$ is the Pauli matrix. Note that $\langle \alpha k|v_y|\alpha k\rangle=0$ is always true for any state.
\begin{figure}
	[t] \centering 
	\includegraphics[width=.4
	\textwidth]{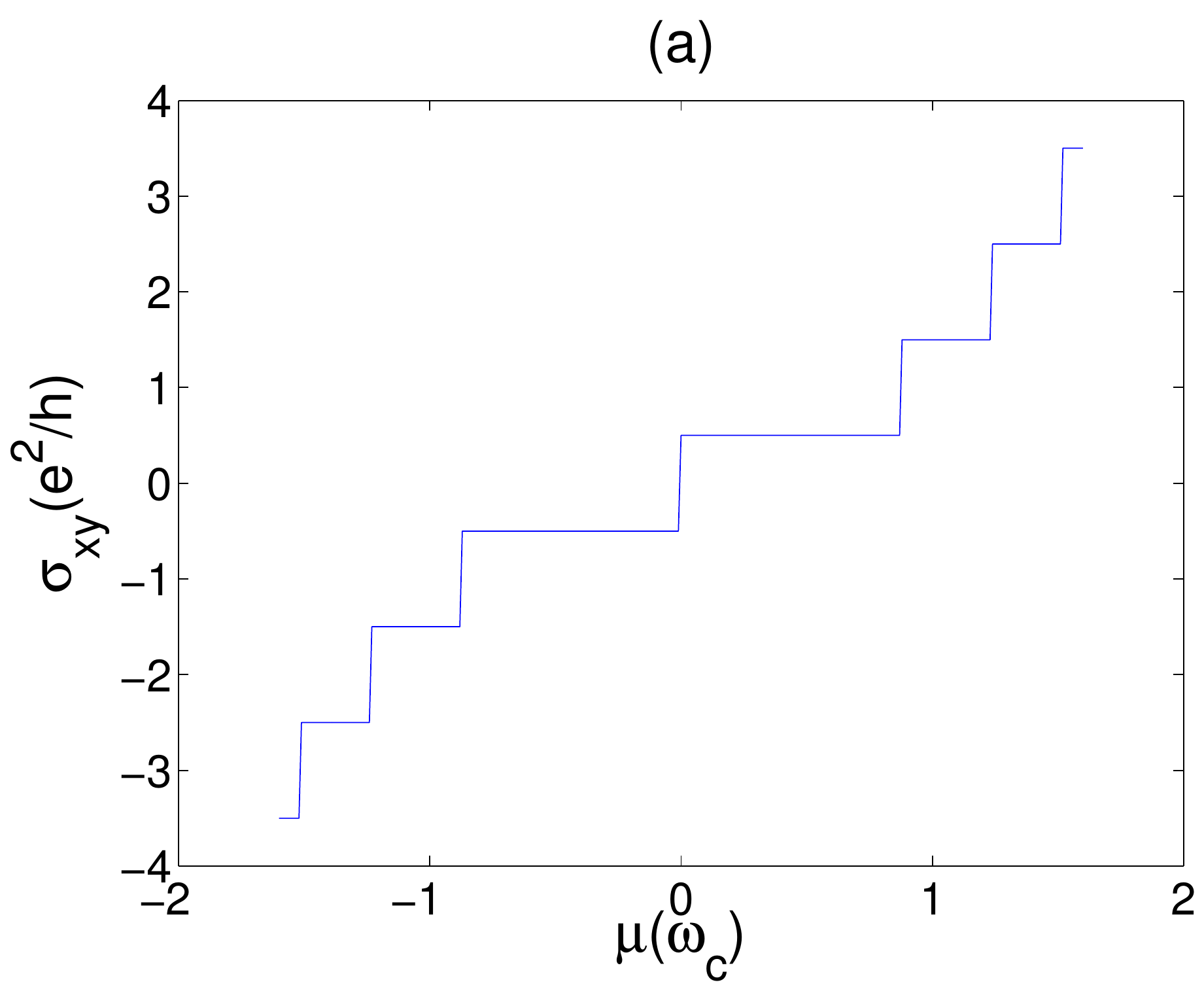} 
	\includegraphics[width=.4
	\textwidth]{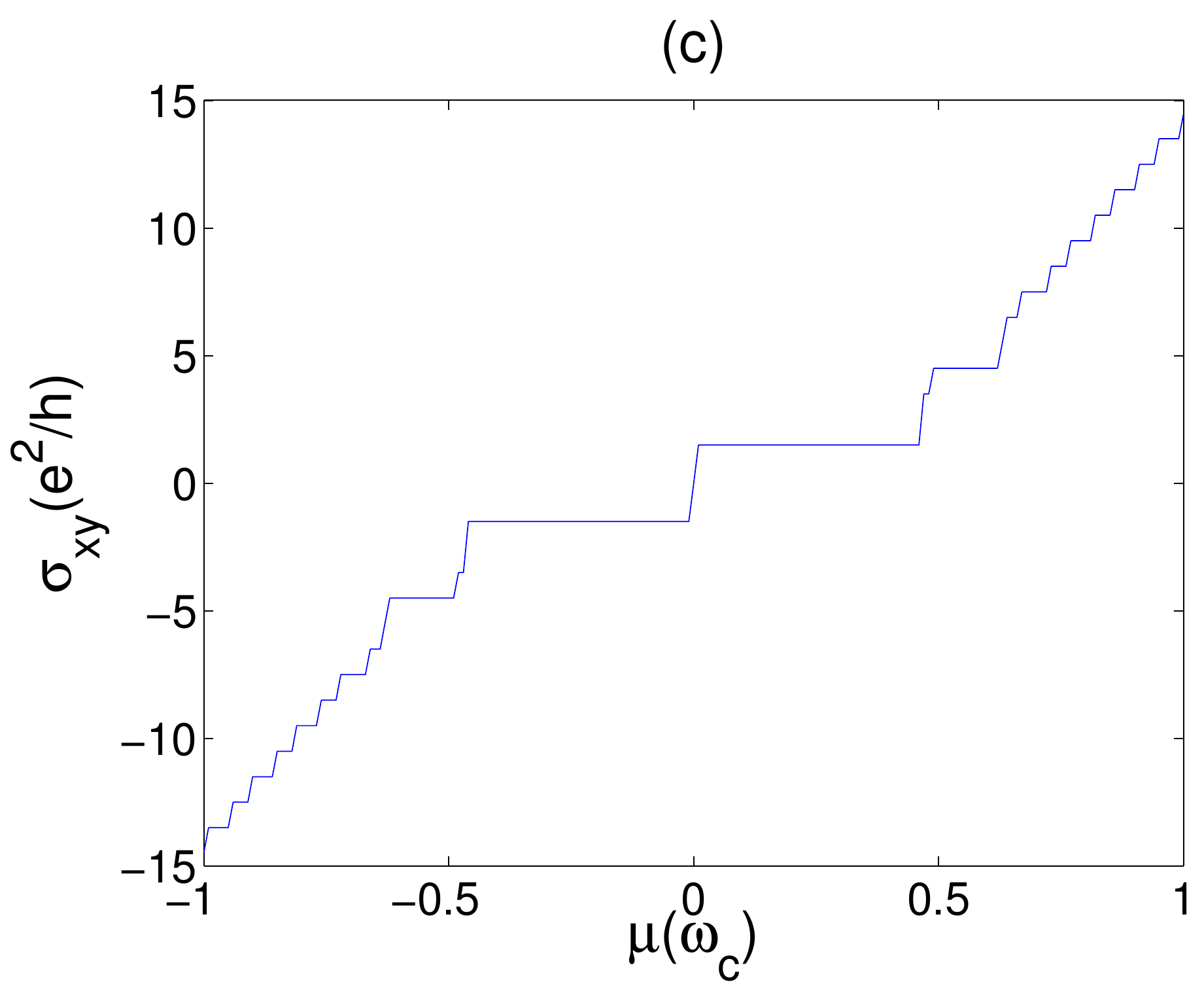} 
	\includegraphics[width=.4
	\textwidth]{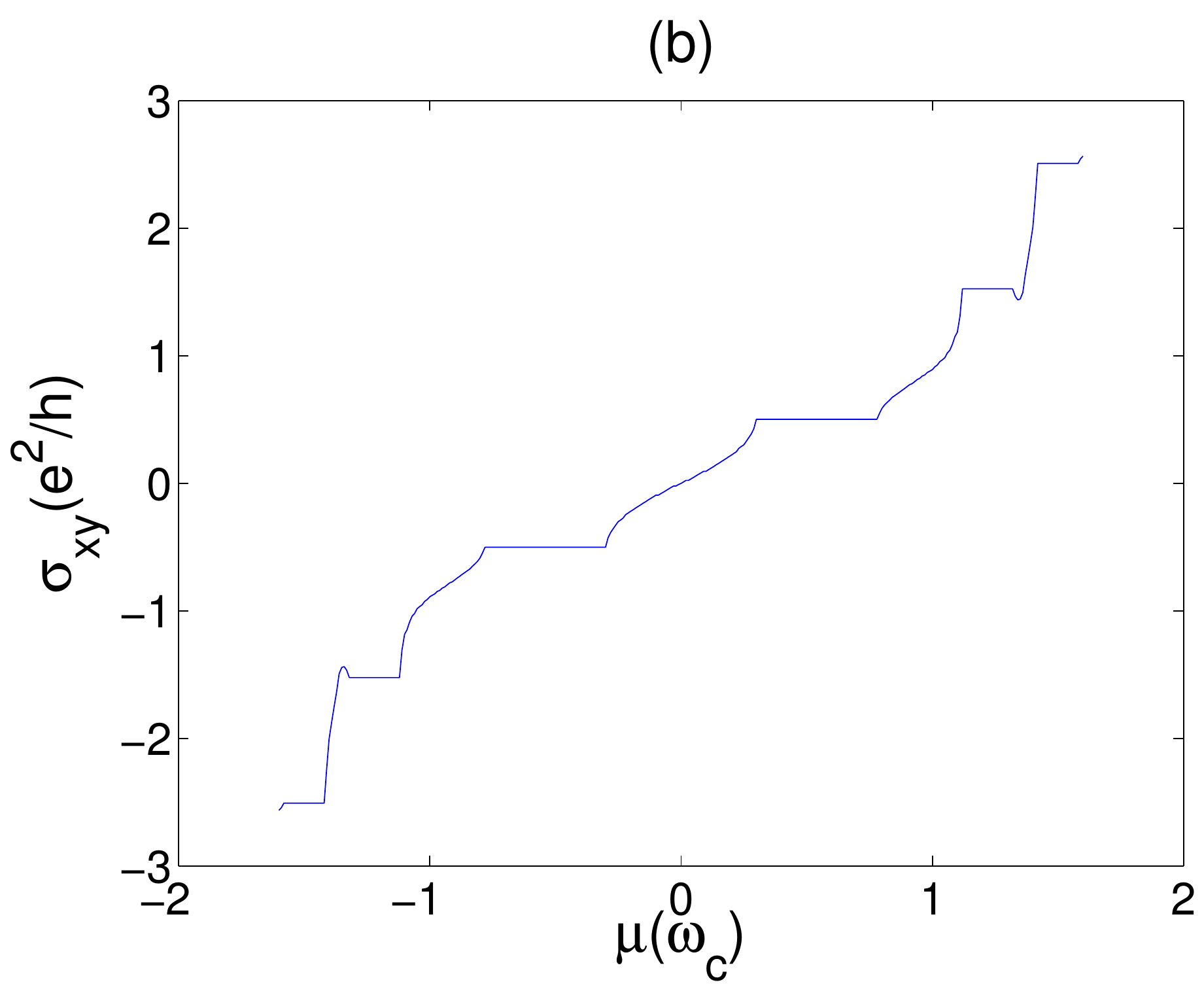} 
	\includegraphics[width=.4
	\textwidth]{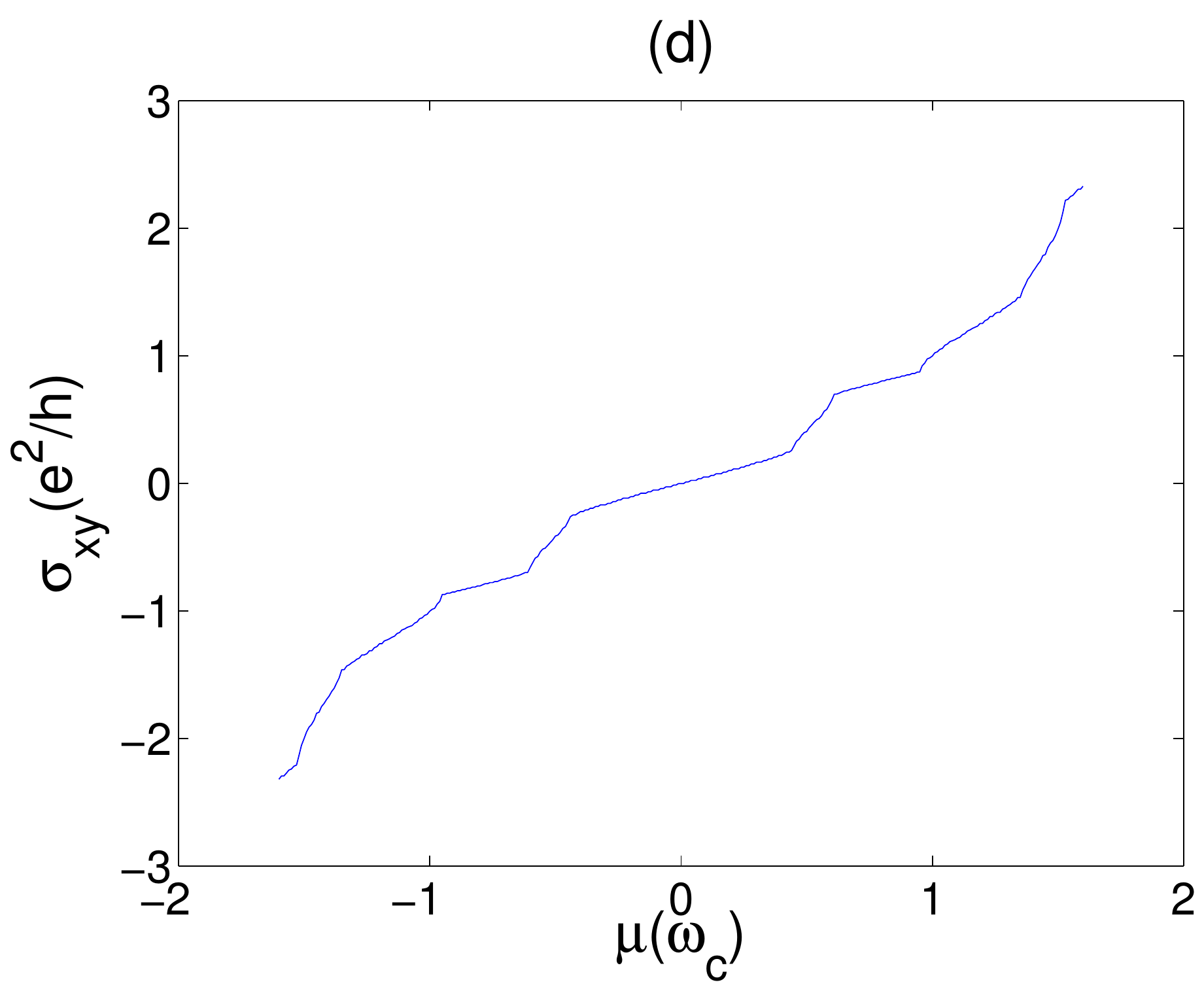} \caption{ (Color online) The dc Hall conductivity of monolayer graphene SL for different (dimensionless) SL strengths $\tilde{U}$, and magnetic fields $B$. The conductivity is shown for weak field ($\ell_B=2\lambda$, top panels) and intermediate field ($\ell_B=0.2\lambda$, bottom panels). Left panels (a,b) correspond to $\tilde{U}=1$, and right panels (c,d) correspond to $\tilde{U}=3$. For weak field (a,c), the Hall conductivity shows well-defined plateaus, as a consequence of nearly flat energy bands. For intermediate field (b,d), the energy bands become dispersive and the Hall conductivity no longer shows step-like structure. However, for weak SL (b), the energy bands are not fully overlapped, Hall conductivity still shows small plateaus when chemical potential falls between two bands, and the value of $\sigma_{xy}$ changes by one between adjacent steps, as expected from Dirac physics. For $\ell_B \ll \lambda$ (not shown), result for pristine graphene is recovered and Hall conductivity is constant between adjacent LLs and changes by one when chemical potential crosses an LL.} \label{Hall_SLG} 
\end{figure}

Results for dc diagonal conductivities as function of chemical potential $\mu$ are shown in Fig. \ref{diagonal_SLG}. This can be done by setting the frequency $\omega$ to zero in Eq. (\ref{Kubo}), and only the real part of the conductivity tensor is nonzero. In weak magnetic field, the conductivities show strong anisotropy, with $\sigma_{yy}$ larger than $\sigma_{xx}$, which is a consequence of the Fermi velocity renormalization in the absence of magnetic field (see Fig. \ref{diagonal_SLG} (a) and (c)). Since $\langle \alpha k|v_y|\alpha k\rangle=0$ and $\langle \alpha k|v_x|\alpha k\rangle\simeq 0$ because of the flat band structure, the major contribution to the diagonal conductivities comes from off-diagonal matrix elements, $\langle \alpha k|v_i|\beta k\rangle$ with $\alpha\neq\beta$. Numerically, we have observed that matrix elements of $v_y$ is always larger than those of $v_x$, which gives rise to the anisotropy in the weak field. In intermediate magnetic field, conductivities still show anisotropy, but with $\sigma_{xx}$ significantly larger than $\sigma_{yy}$ (see Fig. \ref{diagonal_SLG} (b) and (d)). This is because $v_x$ has acquired diagonal matrix element, $\langle \alpha k|v_x|\alpha k\rangle=
\partial E_{\alpha}(y_0=k\ell_B^2)/
\partial k\neq 0$ since the energy spectrum is dispersive, while $v_y$ still lacks this contribution. Notice the positions of the conductivity peaks of $\sigma_{yy}$ exactly correspond to the minimum and maximum of the energy band, where the density of states is the largest. For $\sigma_{xx}$, however, the conductivity is minimum at the band edge, since the average of the velocity operator, $\langle v_x \rangle$, is zero. Therefore, the intra-LL contribution to $\sigma_{xx}$ is the smallest at the band edge. For weak SL potential, $\sigma_{xx}$ can drop to zero when there is no overlapping LLs, while in a strong SL, $\sigma_{xx}$ always show dispersive transport property. In strong magnetic field (not shown here), where the Landau levels structure of pristine graphene is recovered, the conductivities become isotropic (see, for example, Ref. \cite{Ferreira:2011a}). In this case, the SL is merely a perturbation to the magnetic field and thus should have minor effect on determining the magnetotransport properties.

The dc Hall conductivity is shown in Fig. \ref{Hall_SLG}. For weak magnetic field (Fig. \ref{Hall_SLG} (a) and (c)), the Hall conductivity shows well-defined plateaus, as a consequence of nearly flat energy bands. The values of Hall conductivity around Dirac points are $\pm 1/2(e^2/h)$ in weak SL ($\tilde{U}=1$) and $\pm 3/2(e^2/h)$ in strong SL ($\tilde{U}=3$).  This result resembles the anomalous half integer quantum Hall effect in pristine graphene and the Hall conductivity triples due to the existence of three Dirac points in a strong SL. Moving away from the Dirac point, we can observe quantum Hall plateaus with higher conductivities, and the value increases by 1 each time the chemical potential crosses an LL. For intermediate magnetic field (Fig. \ref{Hall_SLG} (b) and (d)), there is no longer well defined plateaus due to the dispersive energy spectrum. However, for weak SL, the LLs are not overlapped with each other. If chemical potential falls between two LLs, a small plateau can still show up, with the value expected from Dirac physics. When magnetic field becomes strong enough as the LL structure for pristine graphene is restored, Hall conductivity will show anomalous half integer quantum Hall plateaus.

\begin{figure}[t] 
	\centering 
	\includegraphics[width=.4\textwidth]{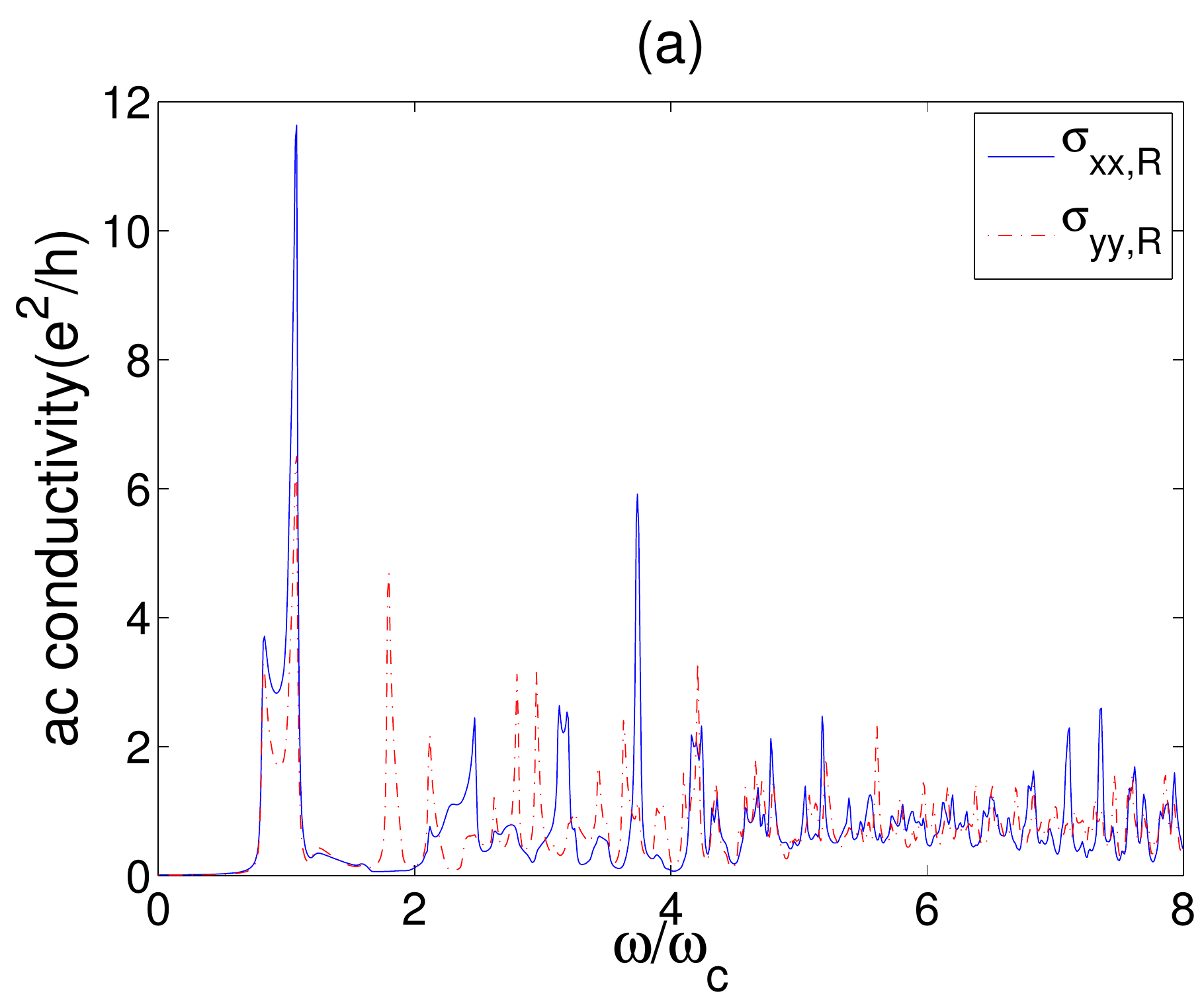} 
	\includegraphics[width=.4\textwidth]{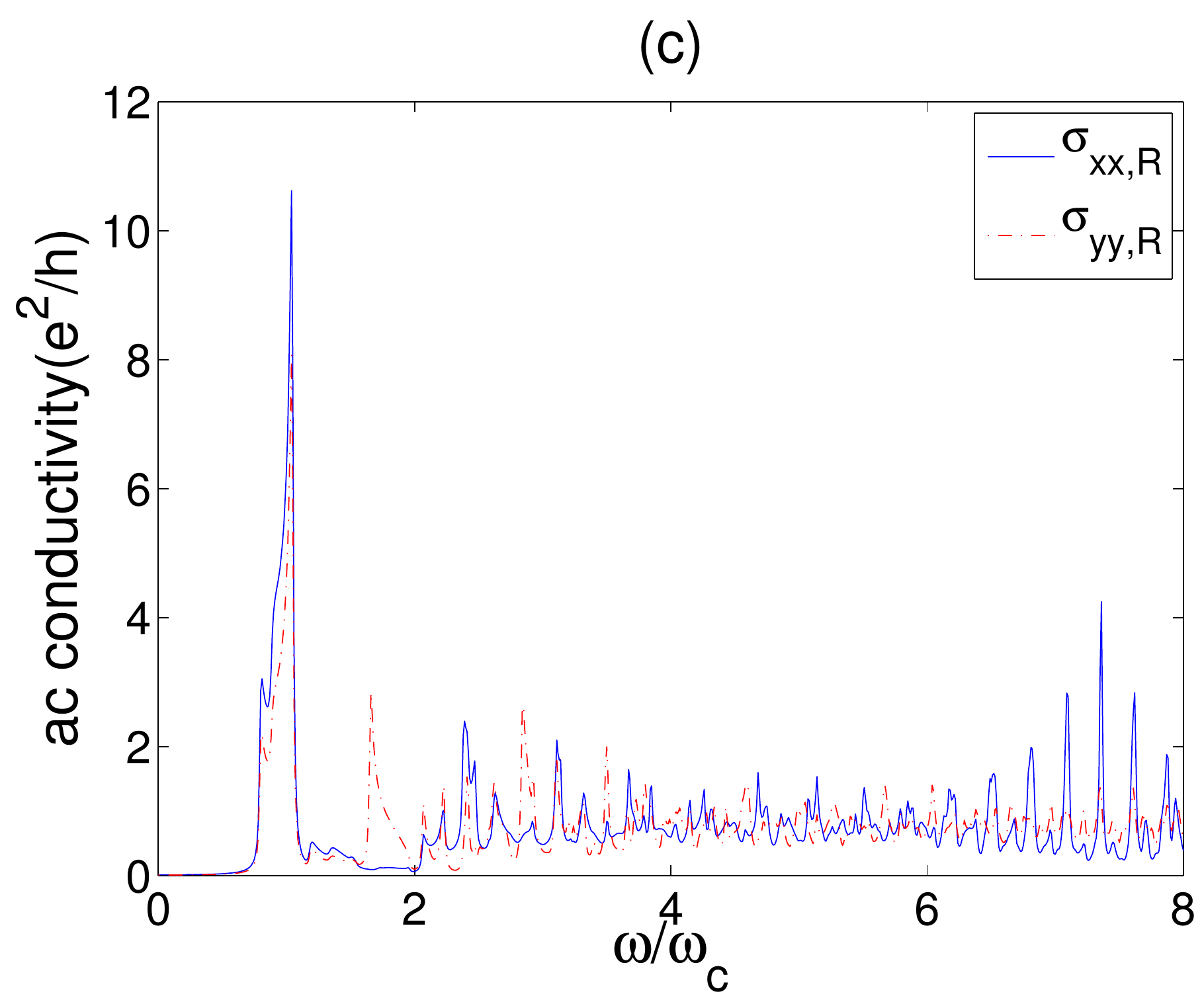} 
	\includegraphics[width=.4\textwidth]{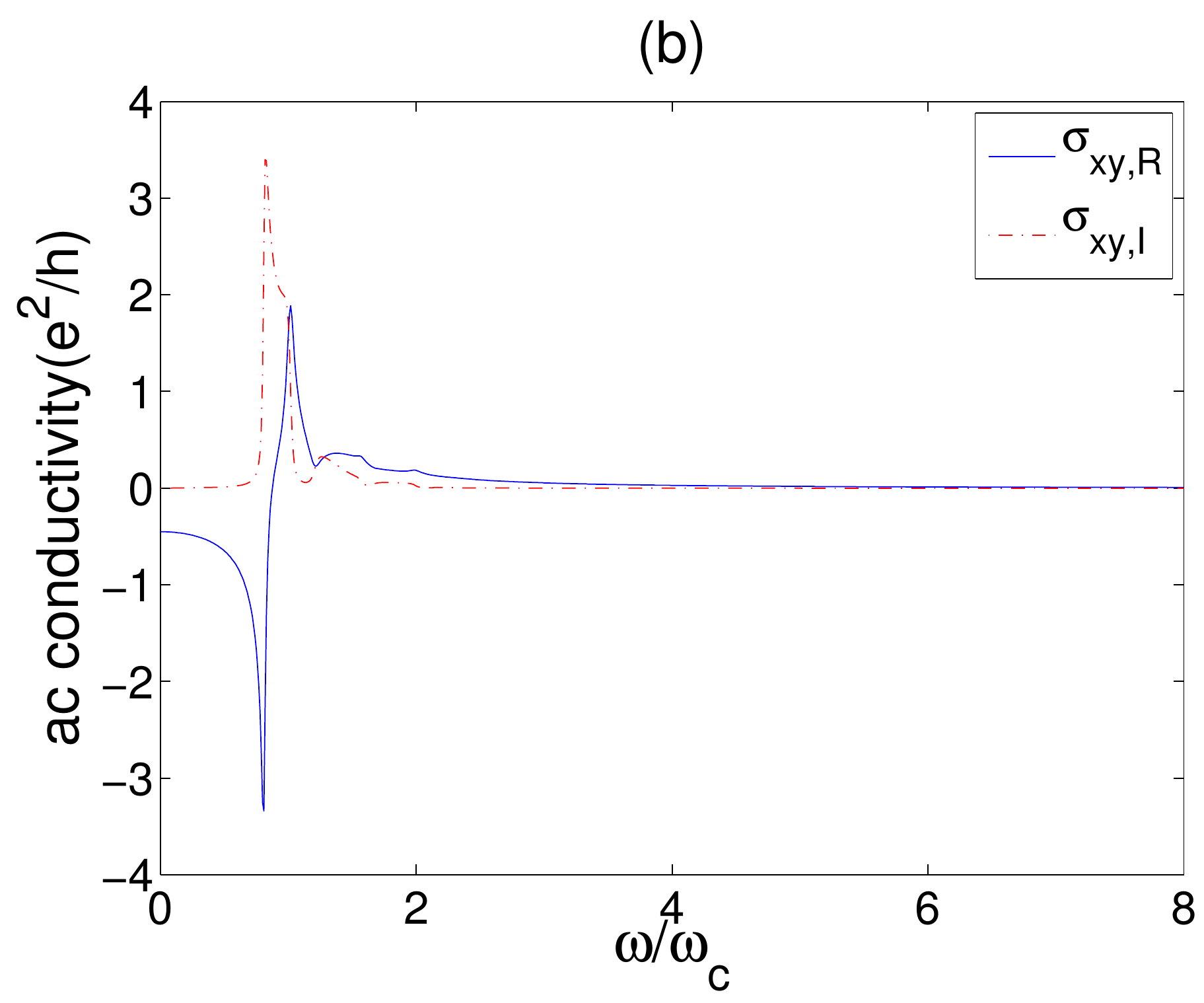} 
	\includegraphics[width=.4\textwidth]{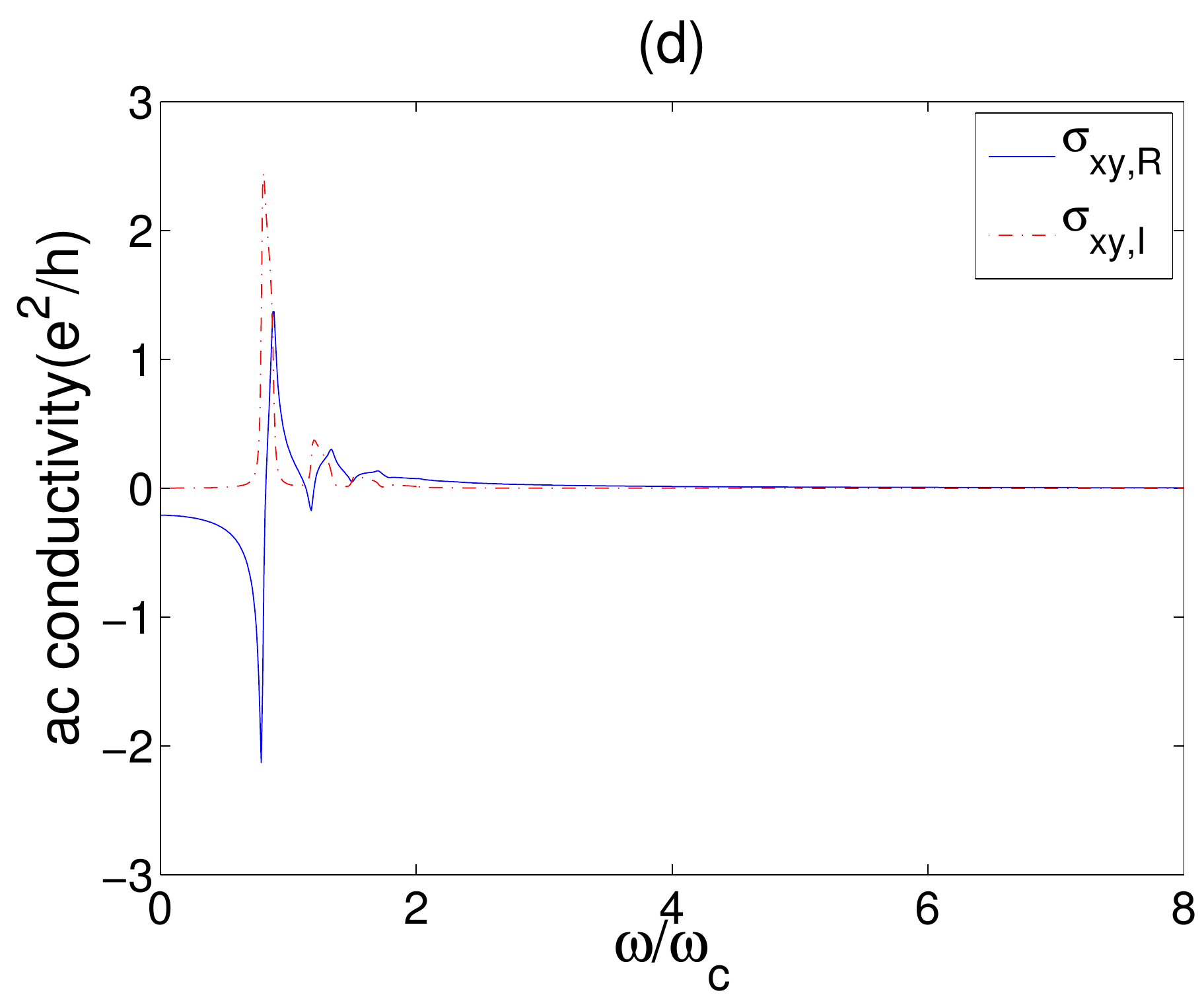} 
	\caption{ (Color online) The ac conductivity of monolayer graphene SL for different (dimensionless) SL strengths $\tilde{U}$, in intermediate magnetic field, $\ell_B=0.2\lambda$, with $\mu=0.2\omega_c$. Left panels (a,b) correspond to $\tilde{U}=1$, and right panels c,d) correspond to $\tilde{U}=3$.} 
	\label{ac_SLG} 
\end{figure}

Fig. \ref{ac_SLG} shows the ac conductivities of graphene SLs in an intermediate magnetic field. For weak and strong 
magnetic fields, the results resemble those of pristine graphene,\cite{Ferreira:2011a} 
since in both cases the LLs are nearly 
flat and the real part of the conductivities show strong peaks when photon frequencies exactly correspond to the energy 
differences between two LLs. In an intermediate magnetic field, the result is complicated by the dispersion of LLs. At 
low frequencies, there can be optical transitions in a range of photon energies, and the real part of diagonal 
conductivities is maximum at the band edge where the DOS is also maximum. At high frequencies, the LLs become less 
dispersive and peaks will show up. These results can be linked with graphene's unusual magneto-optical properties, 
for example, giant Faraday rotation.\cite{Ferreira:2011a,Crassee:2011} 
While the anisotropy in the diagonal conductivities can lead to 
anisotropic rotation angles for incident waves with different polarization plane, this effect is actually quite small 
and hard to observe experimentally.

\subsection{Bandstructure of 2D superlattices}

In 2D SLs, the Fermi velocity near the Dirac point is anisotropically renormalized along every direction. Due to the chiral nature of low energy excitation, there are still energy band crossing at the MBZ boundary.

\begin{figure}[t] 
	\centering 
	\includegraphics[width=.59\textwidth]{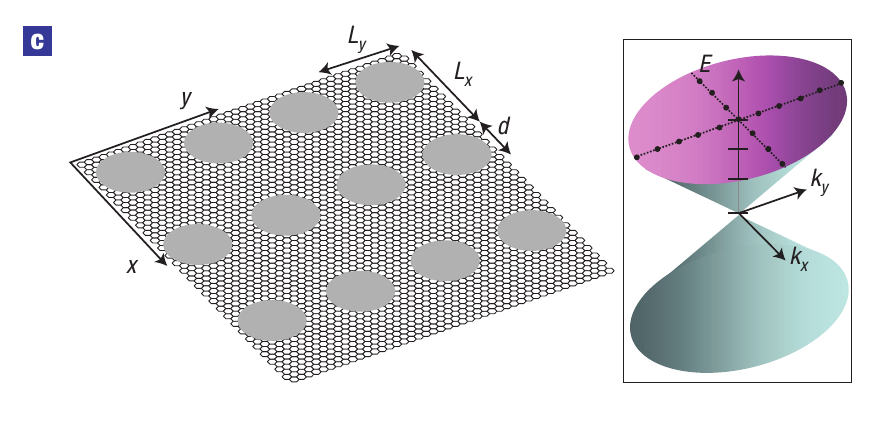} 
	\includegraphics[width=.39\textwidth]{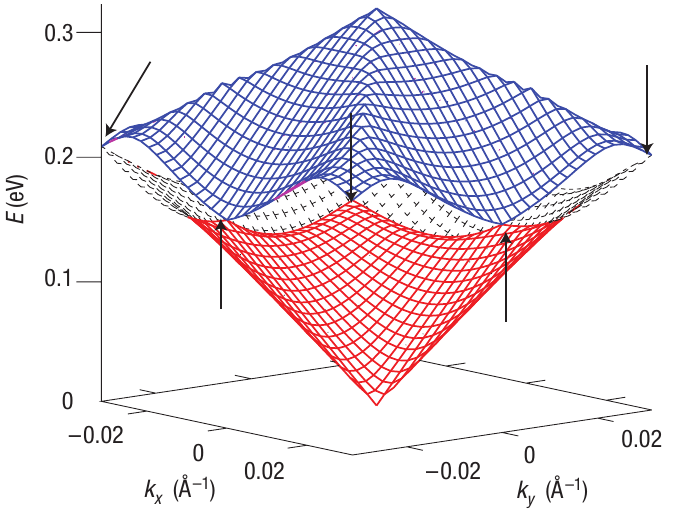} 
	\caption{ (Color online) 2D rectangular muffin-tin type SL potential leads to anisotropically          
	renormalized Dirac cone (left) but no minigap at the MBZ boundary (right).
	[Reprinted by permission from Macmillan Publishers Ltd:
	Nature Physics {\bf 4}, 213, (2008).]
	}
	\label{SquareSL} 
\end{figure}

Fig. \ref{SquareSL} shows a 2D rectangular SL with muffin tin type SL potential with period $L_x$ and $L_y$ in $x$ and $y$ directions, and the corresponding energy spectrum. In contrast to 1D SL where Fermi velocity parallel to the SL direction is not affected, Fermi velocity in a rectangular SL is renormalized in every direction. This can be clearly demonstrated by second order perturbation, assuming a weak SL potential strength,\cite{Park:2008}
\beq
	\frac{v_{\hat{k}}-v_f}{v_f}=-\frac{2\pi^2 U_{{\rm 2D}}^2d^2}{v_f^2L_x^2L_y^2}
	\sum_{\bG\neq 0}\frac{1}{G^4}J_1^2\left(\frac{Gd}{2}\right)\sin^2\theta_{\bk,\bG},
\eeq
where $U_{{\rm 2D}}$ is the 2D SL potential strength in a circular region of diameter $d$, $\bG=(2\pi m/L_x,2\pi n/L_y)$ is the reciprocal lattice vector with $m$ and $n$ integer, and $J_1(x)$ is the Bessel function. Since $\bG$ can be along any direction, compared to 1D SL where $\bG$ is always along the SL direction, we can see that the Fermi velocity is renormalized in every direction. 

The energy spectrum of a 2D rectangular SL also has band crossing points in the middle of an MBZ boundary edge, similar to 1D SL. In addition to these crossing points, at the four corners of the MBZ, energy gap also closes. When similar calculation is carried out for an artificial non-chiral electron, these band crossing points disappear, which truly suggests that they are the consequence of the chirality of low energy excitations in MLG.

\begin{figure}[tb] 
	\centering 
	\includegraphics[width=.39\textwidth]{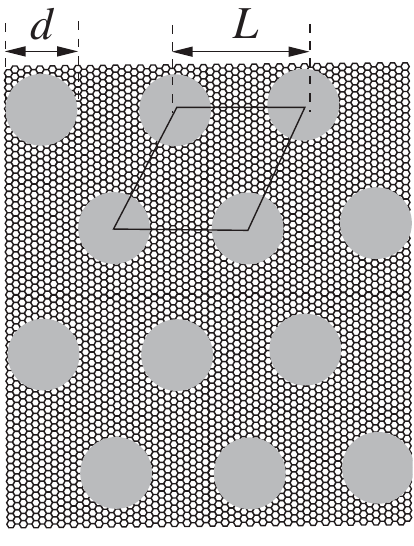} 
	\includegraphics[width=.59\textwidth]{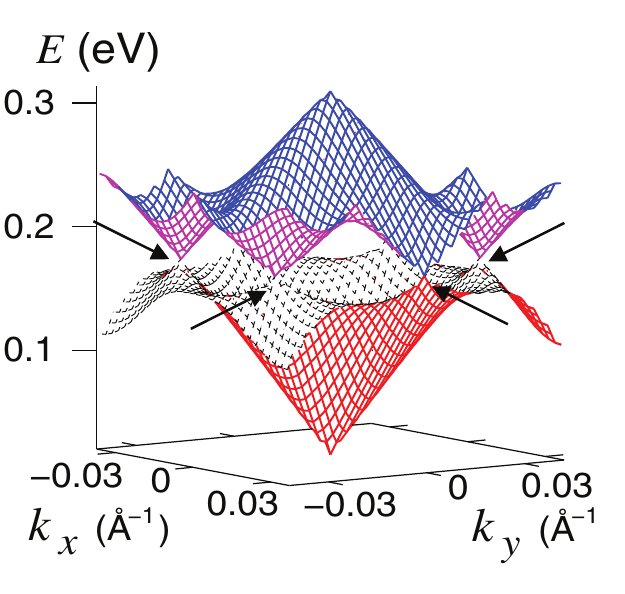} 
	\caption{ (Color online) 2D triangular muffin-tin type SL potential (left) gives rise to finite energy massless Dirac fermions (right).
	[Reprinted with permission from 
	Ref.\cite{Park:2008b}. Copyright (2008) American Physical Society.]
	}
	\label{TriSL} 
\end{figure}

Even though band crossing points appear at the MBZ boundary in both 1D and 2D rectangular SLs, the density of states does not vanish at the crossing energy and the newly generated massless Dirac fermions are obscured by other states. However, for triangular SLs, there exists an energy window where the only available states come from the newly generated massless Dirac fermions.\cite{Park:2008b} Fig. \ref{TriSL} shows a triangular SL with muffin-tin type SL potential and its corresponding energy spectrum. Again, the Fermi velocity is anisotropically renormalized in every direction. The gap between the first and second conduction bands vanishes in the middle of the MBZ boundary edges, and the density of states also vanishes linearly here. This result will have significant impact on the experiments explained later.\cite{Pletikosic:2009,Rusponi:2010,Yankowitz:2012,Ortix:2011}

When graphene is expitaxially grown on a substrate ({\it i.e.}, SiC, hexagonal boron nitride (hBN), transition metal surfaces, etc.), the lattice mismatch between graphene and the substrate and also their relative orientation can lead to a 2D SL with large period. Therefore, theoretical results can be tested on these structures. However, previous results are based on an effective Hamiltonian approach which assumes that external potential does not break sublattice symmetry. When such a symmetry breaking effect is taken into account, most of the earlier results will be modified. For example, a gap should open up at Dirac point and minigap should appear at the MBZ where bands are backfolded. Pletikosi\'{c} et. al.\cite{Pletikosic:2009} have observed a minigap in graphene expitaxially grown on Ir(111) surface, which is due to Moir\'{e} patterned periodic potential. They could not determine whether the Dirac point is gapped because graphene on Ir(111) is slightly $p$-doped. On the other hand, Rusponi et. al.\cite{Rusponi:2010} showed that, in the presence of sublattice symmetry breaking SL potential, the Dirac point remains intact and, remarkably, the Fermi velocities are anisotropically renormalized and the energy spectrum becomes trigonally warped. This is consistent with the theory of Ortix et. al.,\cite{Ortix:2011} where it was demonstrated, incommensurate Moir\'{e} patterned superstructure preserves the Dirac cone in a renormalized form, with threefold global symmetry due to a substrate-induced trigonal warping. Moreover, additional finite energy Dirac points are also generated, but at different positions of MBZ in contrast to Park et. al.\cite{Park:2008b} Since the SL potential also breaks the particle-hole symmetry, the energy spectrum no longer possesses this symmetry, and only in an energy window below the original Dirac point, the newly generated massless fermions are truly Dirac fermions and the density of states can become zero, while for those above, massless fermion states are obscured by the presence of other states. Recently, a scanning tunnelling microscope measurement of graphene on hBN has observed dips in the differential conductance and thus confirmed the existence of finite energy Dirac points.\cite{Yankowitz:2012}

%\section{Superlattices in bilayer graphene}
%\subsection{Band structure of 1D chemical potential superlattices}
%Include local density of states (zero field)?
%\subsection{Band structure of 1D electric field superlattices}
%Include local density of states (zero field)?
%\subsubsection{Kink in the electric field: Topological modes}
%\subsubsection{Interaction effects: Realizing a tuneable 2-band Luttinger liquid}
%\subsection{Landau levels of 1D superlattices}
%\subsection{Magnetotransport in a weak magnetic field}
%\subsection{Band structure of 2D superlattices}

\section{Superlattices in bilayer graphene}

%%%%%%%%%%%%%%%%%%%%%%%%%%%%%%%%%%%%%%%%%%%%%%%%%%%%%%%%%%%%%%%
%%%%%%%%%%%%%%%%%%%%%%%%%%%%%%%%%%%%%%%%%%%%%%%%%%%%%%%%%%%%%%%
%%%%%%%%%%%%%%%%%%%%%%%%%%%%%%%%%%%%%%%%%%%%%%%%%%%%%%%%%%%%%%%
We now turn our attention towards 1D electrostatic potential modulations in BLG. In general, the features of the band structure will depend on the details of the superlattice potential\cite{Barbier:2009,Barbier:2010}. For the bilayer system, a general modulation can be decomposed into two basic types: i) a chemical potential modulation where both layers sit at same potential and, ii) an electric field modulation where there is a local interlayer bias. If the SL potential is purely of one type, there is a dramatic restructuring of the band structure, particularly at low energy. Notably, the low energy quasiparticles transform from being massive chiral fermions in intrinsic BLG to massless chiral Dirac fermions for certain SL parameters. In both cases, much of the band structure can be understood by appealing to the inherent symmetries and/or to an intuitive effective low energy model. The generation of new zero energy modes has similarly been shown to arise in a periodic array of $\delta$-function potentials\cite{Barbier:2010b}, twisted BLG \cite{Gail:2011}, and along domain walls in monolayer graphene with broken sublattice symmetry \cite{Semenoff:2008,Jung:2012}. 

In this section, we focus on reviewing the band structure of the two rudimentary types of SL in BLG, a chemical potential and electric field superlattice. Both types of SL are of particular interest because each can support the formation of new Dirac points for arbitrarily \textit{weak} SL strengths, in contrast to SL in the monolayer. In fact, the Dirac points for the electric field SL survive even for strong modulations. A thorough understanding of these two basic SL potentials also provides a firm foundation to understand more generic SL profiles and the formalism reviewed here can readably be applied to more general SLs.

We start here by introducing the low energy Hamiltonian that can be used to study the properties of generic SL of moderate strength. It should be noted that for larger SL potentials, the full tight-binding model is required to correctly describe new features in the band structure. Instances where the full Hamiltonian gives quantitative differences in the band structure will be duly noted. After establishing the formalism, we discuss the band structure generated by a chemical potential and electric field SL in Sec.~\ref{sect:BLG_chem} and Sec.~\ref{sect:BLG_electric}, respectively.

The low energy Hamiltonian for Bernal-stacked BLG can be obtained by expanding its minimal tight binding spectrum near one of the Brillouin zone corners ($\bK$ points).\cite{McCann:2006} When the layer potential (i.e., interlayer potential difference) is not too large, $|\Delta|\ll t_{\perp}$, we find ${\mathcal H}=\psi^{\dagger}\hat{H}\psi$,\cite{McCann:2006} where 

\begin{equation}
	\label{Hred} \hat{H}=-\frac{v_F^2}{t_{\perp}}\left( 
	\begin{array}{cc}
		0 & (s p_x + ip_y)^2 \\
		(s p_x - ip_y)^2 & 0 
	\end{array}
	\right)+\left( 
	\begin{array}{cc}
		V_1(x,y) & 0 \\
		0 & V_2(x,y) 
	\end{array}
	\right), 
\end{equation}
and $\psi^{T}=(a_{\bf x},b_{\bf x})$, with $a$ ($b$) being the electron operator on the top (bottom) layer. Here, $p_{x(y)}=-i\partial_{x(y)}$ is the momentum operator, $s=\pm 1$ for the Hamiltonian at the $\pm \bK$ valley, $v_F \! =\! \sqrt{3}td/2 \! \approx \! 10^6$~m/s is the Fermi velocity, $t \! \approx \! 3$~eV is the nearest neighbor hopping integral, $d \! \approx \! 2.46$ \AA \, is the distance between neighboring atoms on the same sublattice (note: $d=a \sqrt{3}$ where $a$ is the nearest neighbor
Carbon-Carbon distance), $V_{1,2}$ are the potentials on each layer, and $t_{\perp} \!\approx \! 0.15 t$ is the interlayer coupling. Unless stated, we set $t\!\!=\!\!d\!\!=\!\!1$. We will ignore inter-valley scattering assuming the potentials are varying slowly on the scale of $d$, so we only consider the $s=+1$ valley (at $\bK$). Such an approach has been successfully used to study SLs in monolayer graphene \cite{Park:2008,Park:2008a}.

To diagonalize $H_{\rm kin}$, we Fourier transform and then make a unitary transformation $a_{\bf p} \!=\! (\alpha_{\bf p} \!+\! \beta_{\bf p})/\sqrt{2}$, $b_{\bf p} \!=\! {\rm e}^{2 i \theta_{\bf p}} (\alpha_{\bf p} \!-\! \beta_{\bf p})/\sqrt{2}$, where $\cos\theta_{{\bf p}}\!=\!p_x/p$ and $p\!=\!\sqrt{p_x^2+p_y^2}$. This leads to $H_{\rm kin}\!=\!\sum_{\bf p}\left(\varepsilon_{e}({\bf p})\beta^{\dagger}_{\bf p} \beta^\pdg_{\bf p} \!+\! \varepsilon_{h}({\bf p})\alpha^{\dagger}_{\bf p}\alpha^\pdg_{\bf p}\right)$. Here $\varepsilon_{e,h}({\bf p})\!=\!\pm p^2/2 m^*$ are energies of electron (hole) states, with an effective mass $m^* \!\equiv\! t_{\perp}/(2 v_F^2)$. This minimal model supports quadratic band touching points at $\pm \bK$.

When $V_{1,2}({\bf x})$ are periodic, we can also Fourier transform the SL potential to obtain $H_{\rm SL}=\sum_{{\bf p},{\bf G}}\Psi^{\dagger}({\bf p}) W_{{\bf p},{\bf G}} \Psi({\bf p}-{\bf G})$, where 
\begin{eqnarray}
	\label{sl} W_{{\bf p},{\bf G}}\!\!=\!\!\frac{1}{2} \left( 
	\begin{array}{cc}
		\! V_1({\bf G})\!+\!V_2({\bf G}){\rm e}^{2i\theta} & V_1({\bf G})\!-\!V_2({\bf G}){\rm e}^{2i\theta} \! \\
		\! V_1({\bf G})\!-\!V_2({\bf G}){\rm e}^{2i\theta} & V_1({\bf G})\!+\!V_2({\bf G}){\rm e}^{2i\theta} \! 
	\end{array}
	\right)\!\!, \label{W} 
\end{eqnarray}
$\Psi^\dg({\bf p})\!=\!(\alpha^\dg_{\bf p},\beta^\dg_{\bf p})$, and $\theta \!\equiv\! \theta_{{\bf p}-{\bf G}}\!-\!\theta_{{\bf p}}$ is the angle between momenta ${\bf p}\!-\!{\bf G}$ and ${\bf p}$. Our aim is to understand the band structures of SLs described by $H_{\rm kin}+H_{\rm SL}$. We will study 1D SLs with period $\lambda$ along $\hat{y}$, so that the reciprocal lattice vectors, $\{\bG\}$, are integer multiples of $\bQ=(0,2\pi/\lambda)$, and the mini Brillouin zone (MBZ) boundaries are at $p_y=\pm \pi/\lambda$.

%%%%%%%%%%%%%%%%%%%%%%%%%%%%%%%%%%%%%%%%%%%%%%%%%%%%%%%%%%%%%%%
%%%%%%%%%%%%%%%%%%%%%%%%%%%%%%%%%%%%%%%%%%%%%%%%%%%%%%%%%%%%%%%
\subsection{Band structure of 1D chemical potential superlattices}\label{sect:BLG_chem}
\begin{figure}
	[th] \centering 
	\includegraphics[width=.38 
	\textwidth]{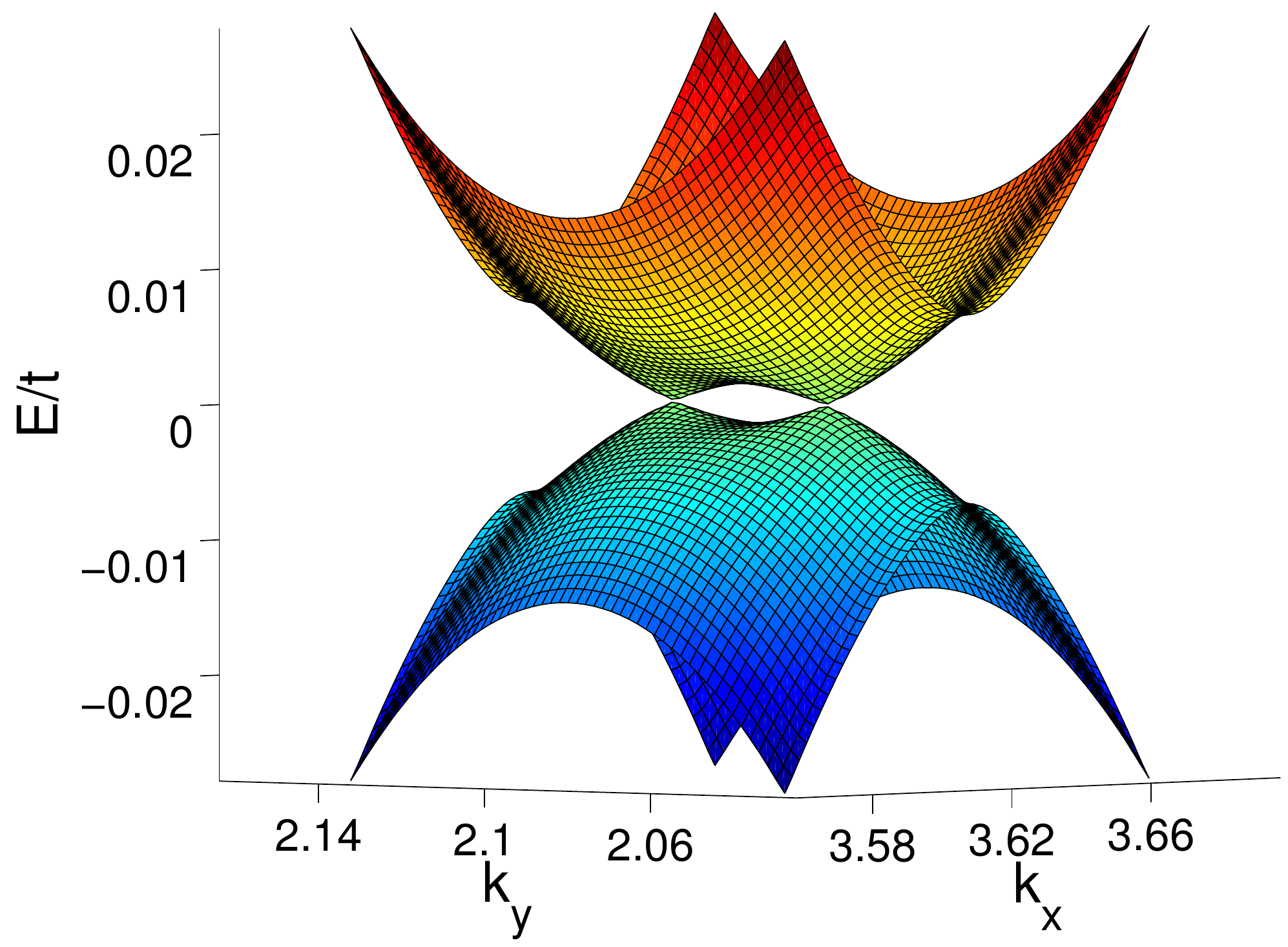} \quad
	\includegraphics[width=.38
	\textwidth]{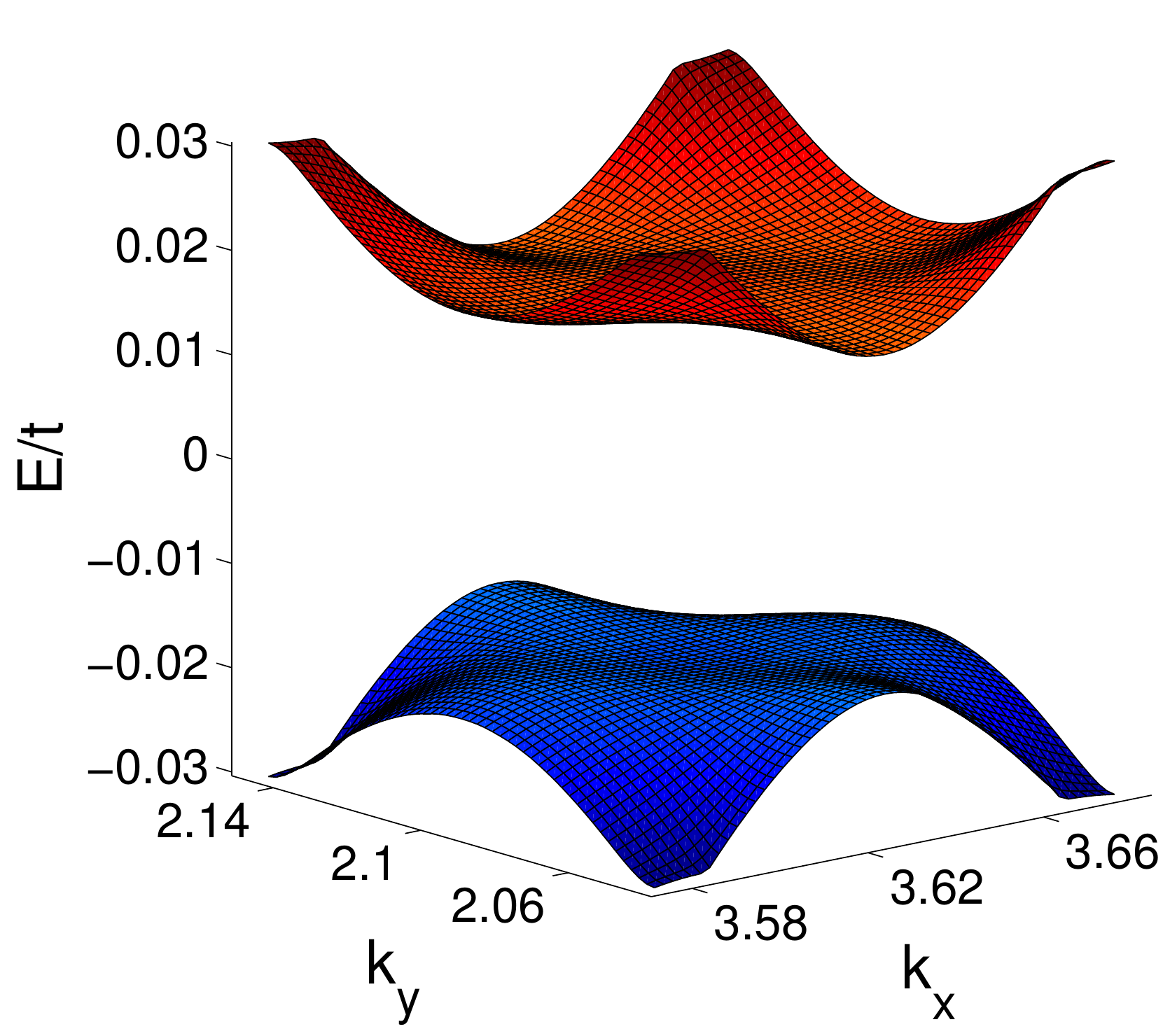} \caption{Energy spectrum for a 1D superlattice with step-like chemical potential modulation of amplitude $U$. We set $\lambda=60d$, with [left panel] $U=0.01t$ showing two Dirac nodes split along $\hat{y}$ near $\bK$, and with [right panel] $U=0.04t$ showing a full gap.} \label{fig1} 
\end{figure}

%%%%%%%%%%%%%%%%%%%%%%%%%%%%%%%%%%%%%%%%%%%%%%%%%%%%%%%%%%%%%%%
%%%%%%%%%%%%%%%%%%%%%%%%%%%%%%%%%%%%%%%%%%%%%%%%%%%%%%%%%%%%%%%
A chemical potential SL corresponds to the case where $V_{1}(x,y)=V_2(x,y)=U(x,y)$. For simplicity, we first consider a step-like potential with with (i) $U(x,y)=U$ for $0 \leq y < \lambda/2$ and (ii) $U(x,y)=-U$ for $\lambda/2 \leq y < \lambda$ and use the effective two-band Hamiltonian introduced above. Starting from $U=0$ and increasing the SL strength to moderate values, we observe the following restructuring of the band dispersion (see Fig.~\ref{fig1}): i) the zero energy quadratic band touching point splits into two anisotropic Dirac cones located at $(0,\pm p^*_y)$, ii) further increasing $U$ causes the Dirac points to push out towards the MBZ and, iii) upon reach the boundary at $(0, \pm \pi/\lambda)$, a band gap opens at a critical $U=U_c$.\cite{Barbier:2010,Tan:2011,Killi:2011a} Before considering the band structure for $U$ beyond $U_c$, let us first discuss the formation of the Dirac cones in more detail. 

\subsubsection{Dirac Cones: Formation} 
The sequence of semimetal to band insulator with increasing SL strength was shown to not be dependent on any symmetry in SL profile and to be robust even against weak perturbations that vary slowly perpendicular to the principle SL direction.\cite{Tan:2011} Given the persistence of the Dirac point, it important to understand why it forms and how it is protected.

The formation of linear band crossing points has been argued to be deeply rooted in the chiral nature of the low energy BLG quasiparticles.\cite{Tan:2011,Killi:2011a} This can be seen from the scattering angle dependence of the matrix elements in Eqn.~\ref{sl} that arises from the pseudospin structure of the eigenstates. For states with momenta parallel to the modulation direction, $\theta=0$ or $\pi$, the off-diagonal matrix elements vanish; the electron and hole states decouple, so that a particle in an electron (hole) state can only forward/back-scattering of the SL potential to another electron (hole) state. Since all such electron (hole) states within the first MBZ in an extended zone scheme only mix with electron (hole) states of higher (lower) energy, the energy of the conduction (valence) band will be globally shifted down (up). This results in two level crossings along the modulation direction, which are protected by the chirality of the low energy BLG quasiparticles. If this electron-hole decoupling was true for all momenta, we would see the two parabolic bands crossing on a full circle in the MBZ, but going to momenta $(\delta p_x, p_y)$ leads to electron-hole mixing that is linear in $\delta p_x$; this results in an avoided level crossing and the robust emergence of two Dirac cones in the MBZ.

\subsubsection{Dirac Cones: Properties} 
The location and velocity anisotropy of Dirac cones, as well as the critical modulation amplitude to gap them out, can be estimated using perturbation theory in $U(\bG)$. The second order energy correction of states with ${\bf p}=(0,p_y)$ is $ \Delta E^{(2)}({\bf p})=\sum_{n\neq 0}{|U(n{\bf Q})|^2}/\left[ {\varepsilon_{e,h}({\bf p})-\varepsilon_{e,h}({\bf p}+n{\bf Q})}\right]. $ Since ${\varepsilon_{e}({\bf p})<\varepsilon_{e}({\bf p}+n{\bf Q})}$ while ${\varepsilon_{h}({\bf p})>\varepsilon_{h}({\bf p}+n{\bf Q})}$ in the MBZ, this correction is always negative (positive) for electron (hole) states, as expected.

Thus, the two bands will intersect and cross linearly at momenta $(0,\pm p^*_y)$, where $ {p^{*2}_y}/{2m^*}=2m^* \sum_{n \neq 0} |U(n{\bf Q})|^2 /\left[{n^2Q^2+2p^*_y nQ}\right]. $ For weak modulations, $p^*_y/Q \! \ll\! 1$, and keeping only $n\!=\!\pm 1$, we estimate $p^*_y \! \approx \! \sqrt{2}m^* |U(\bQ)| \lambda/\pi$. For a step profile, $|U(\bQ)| \!=\! 2 U/\pi$, and $|n| \! > \! 1$ contributions are small. 

For small $\delta p_x$ away from the level crossing point, we can estimate the electron-hole mixing term using perturbation theory and we find that the resulting eigenstates have energies $\epsilon_\bp= \pm (16 m^* |U(\bQ)|^2 / |\bQ|^2) \delta p_x/p^*_y$. The crossing points at $(0,\pm p^*_y)$ are thus really massless Dirac points in the full MBZ. We find velocities $v_y = p^*_y/m^* \approx \sqrt{2} \lambda |U(\bQ)|/\pi$, and $v_x = 2 v_y$ for the anisotropic linear dispersion.

Once these Dirac nodes reach the MBZ boundary, Bragg scattering between them opens up a full gap. The critical potential strength, $|U_c(\bQ)|$ for this is roughly estimated by setting $p_y^*=Q/2$, which yields $|U_c(\bQ)| \approx \pi^2/(\sqrt{2}m^*\lambda^2)$. For a step profile, with $\lambda=60d$, we find $U_c \approx 0.03 t$ which is close to the numerical result $0.02t$.

\subsubsection{Strong Potentials and Electron Screening} 
For larger SL potentials beyond $U_c$, it becomes necessary to employ the four band model to capture a the effects of the high energy bands. Increasing $U$ above $U_c$, the band gap continues to grow until a maximum value is reached and then begins to decrease before finally closing by forming two new pairs of Dirac cones. For even greater $U$, two of the four Dirac cones merge at $(0,0)$ and become gapped, but remaining two Dirac cones retain the semi-metallicity of the system.\cite{Tan:2011,Killi:2011a}

Tan et al.\cite{Tan:2011} also performed a self-consistent tight-binding calculation to describe the higher energy effects of the entire band structure and to determine the effects of interactions on the band dispersion. Besides confirming that the simple single particle low energy model correctly describes the qualitative features of the band structure, it showed that it is possible to account for screening at the Hartree level by a dielectric constant $\epsilon \sim 11$. Hence, the main effect of electron interactions is to screen the external SL potential, therefore increasing the critical SL potential required to open a band gap. 

%%%%%%%%%%%%%%%%%%%%%%%%%%%%%%%%%%%%%%%%%%%%%%%%%%%%%%%%%%%%%%%
%%%%%%%%%%%%%%%%%%%%%%%%%%%%%%%%%%%%%%%%%%%%%%%%%%%%%%%%%%%%%%%
\subsection{Band structure of 1D electric field superlattices}
%%%%%%%%%%%%%%%%%%%%%%%%%%%%%%%%%%%%%%%%%%%%%%%%%%%%%%%%%%%%%%%
%%%%%%%%%%%%%%%%%%%%%%%%%%%%%%%%%%%%%%%%%%%%%%%%%%%%%%%%%%%%%%%
\label{sect:BLG_electric} When BLG is subjected to an electric field SL the potentials on the two layers are such that $V_1(x,y)=-V_2(x,y)=U(x,y)$. In contrast to the chemical potential SL discussed above, the band structure is sensitive to the form of the SL profile\cite{Barbier:2009}. To illustrate this, Killi et al.\cite{Killi:2011a} considered a more general periodic potential, with $U(y)=2 U (1-w/\lambda)$ for $0 \leq y < w$, and $U(y)=-2 U w/\lambda$ for $w \leq y < \lambda$, where we have kept the average potential to zero. 

\begin{figure}
	[t] \centering 
	\includegraphics[width=.35 
	\textwidth]{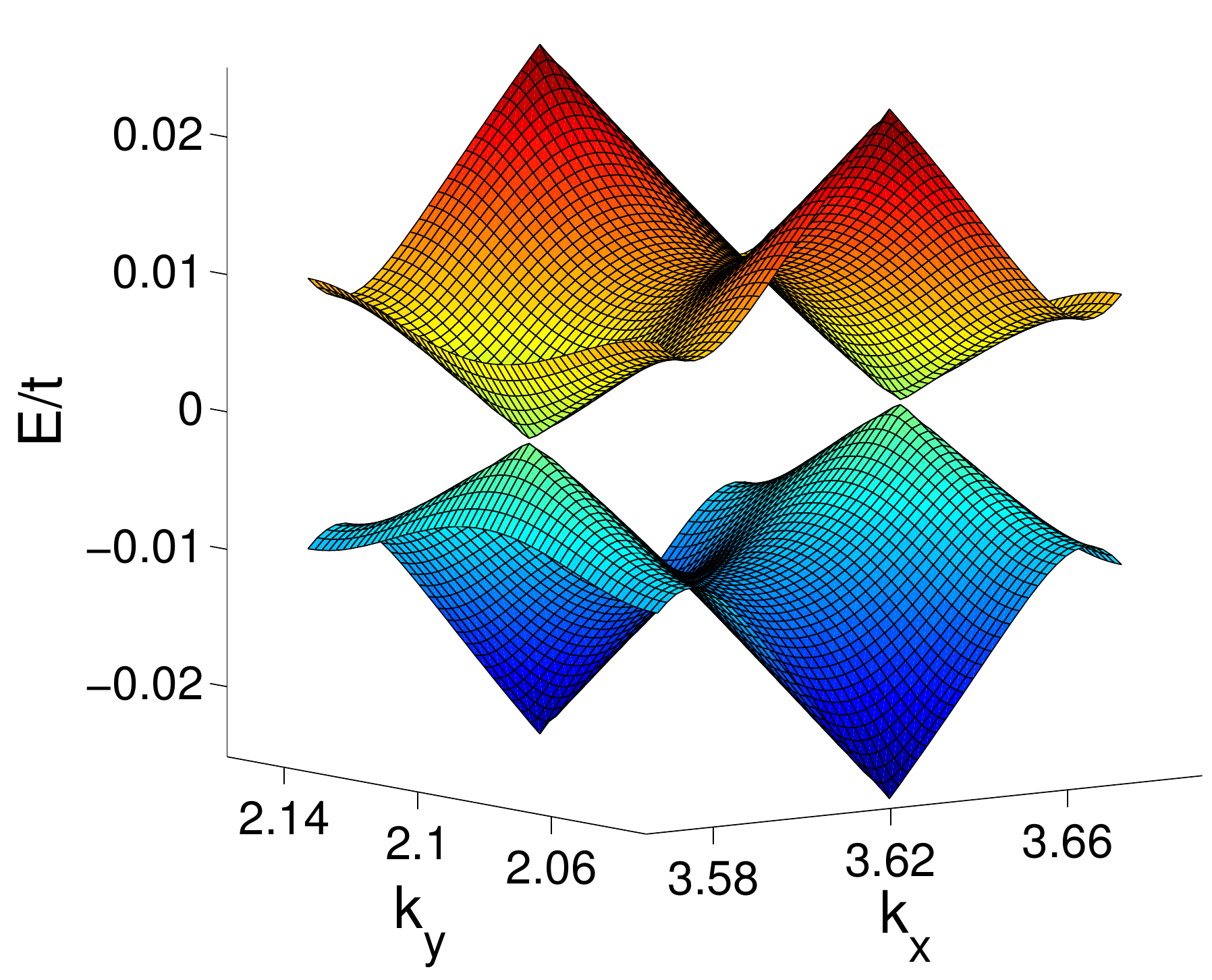} \quad
	\includegraphics[width=.35 
	\textwidth]{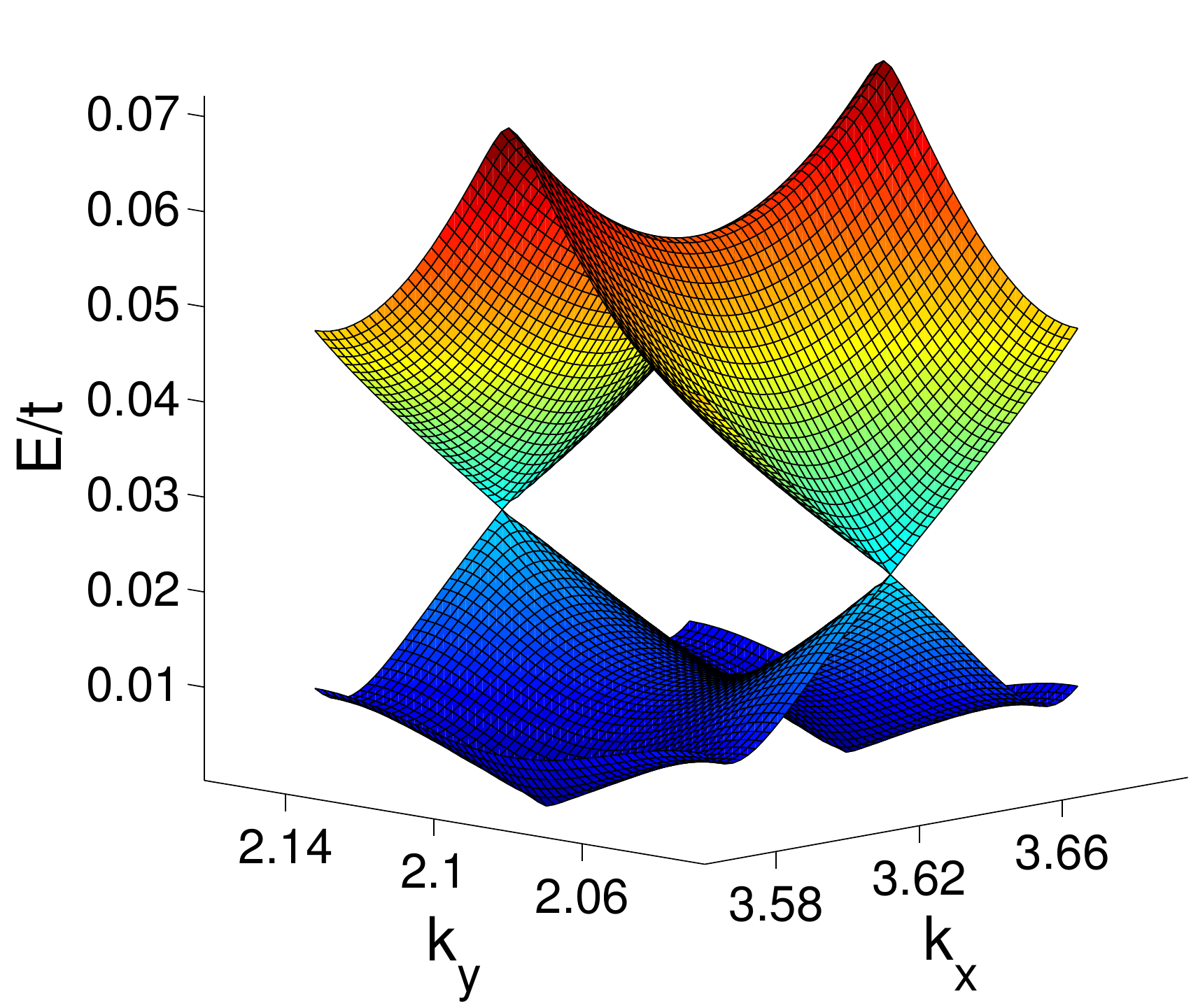} \caption{Energy spectrum for a 1D {\it symmetric} (see text) electric field superlattice with $\lambda=60d$ and $U=0.03 t$, showing a pair of zero energy massless Dirac fermions at $(\pm p^*_x,0)$ [left panel] and a nonzero energy Dirac point at $(0,\pm \pi/\lambda)$ [right panel].} \label{fig3} 
\end{figure}

If the parameter $w=\lambda/2$ the SL potential is {\it symmetric}. A numerical calculation of the band structure show a pair of anisotropic massless Dirac cones forming at zero energy at $(\pm p^*_x, 0)$, as seen in Fig.~\ref{fig3} (left panel)\cite{Barbier:2010a,Killi:2011a}. Here, the zero energy Dirac cones lie along the direction perpendicular to the modulation as opposed to along it for chemical potential SL. Two additional anisotropic Dirac cones are also present at high energy, one in the valence band the other in the conduction band. Irrespective of the strength of the SL potential, these Dirac points are pinned MBZ boundary at $(0,\pi/\lambda)$ (or equivalently $(0,-\pi/\lambda)$).

When $w \neq \lambda/2$ a band gap opens at all of the Dirac points. This suggests that the protection of the Dirac points is governed by a symmetry of the SL profile that is broken when $w \neq \lambda/2$. It is found that the relevant symmetry corresponds to a generalized parity operator ${\cal P}$ that transforms $y \to -y$ followed by exchanging the two layers of BLG, and the Dirac points persist as long $[{\cal P}, H]=0$. 

In the letter by the present authors,\cite{Killi:2011a} a simple intuitive picture for understanding all of the features of the Dirac cones was proposed. The idea is to view the SL potential as establishing a periodic array of `kink' and `anti-kink' steps in the potential profile where the parity of interlayer bias reverses. Since it was shown by Martin et al.\cite{Martin:2008} that topological zero energy modes are confined along an isolated kink (or anti-kink), it is possible to construct an effective low energy theory evolving these modes. Moreover, the band structure of the SL should be entirely dictated by these modes fore energies below $U$ where the bulk states are gapped. As a prerequisite to presenting and analyzing the low energy effective model, it is necessary to understand some of the basic properties of the 1D kink modes. The next we provide a brief overview of these states before returning the construction of the low energy model of the electric field SL.

%%%%%%%%%%%%%%%%%%%%%%%%%%%%%%%%%%%%%%%%%%%%%%%%%%%%%%%%%%%%%%%
\subsubsection{Kink in the electric field: Soliton modes}
%%%%%%%%%%%%%%%%%%%%%%%%%%%%%%%%%%%%%%%%%%%%%%%%%%%%%%%%%%%%%%%
\label{sect:Kink}
For a general potential profile with $V_g(y>0)=-V_g(y<0)$ and $V_g( y\to\pm\infty) = \pm V_g$, the bulk region far from $y=0$ has a gap $\Delta \approx V_g$, while along this interface, localized `topological' edge modes emerge that are analogous to those in quantum hall systems.\cite{Martin:2008}  These modes can be thought of as forming chiral 1D quantum wires, since states from opposite valleys are counterpropagating (this follows from the Berry curvature about each valley having opposite sign).  With respect to the two band Hamiltonian, the kink interface marks a region where the mass of the quasiparticles, $\propto \sigma_z$, changes sign.

\begin{figure}[h]  \centering 
	\includegraphics[height=2.8cm]{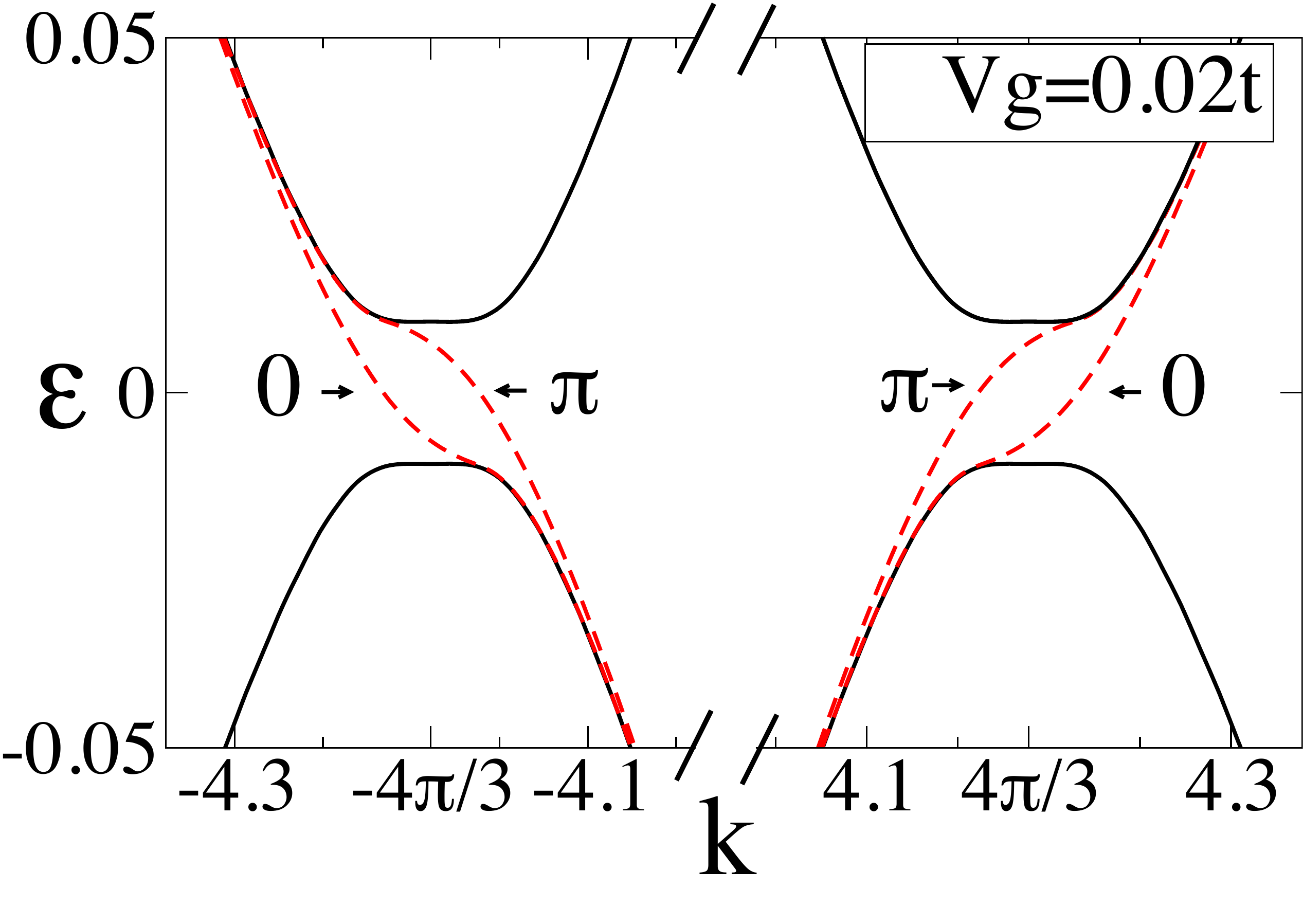} \hspace{-.25cm} 
	\includegraphics[height=2.8cm]{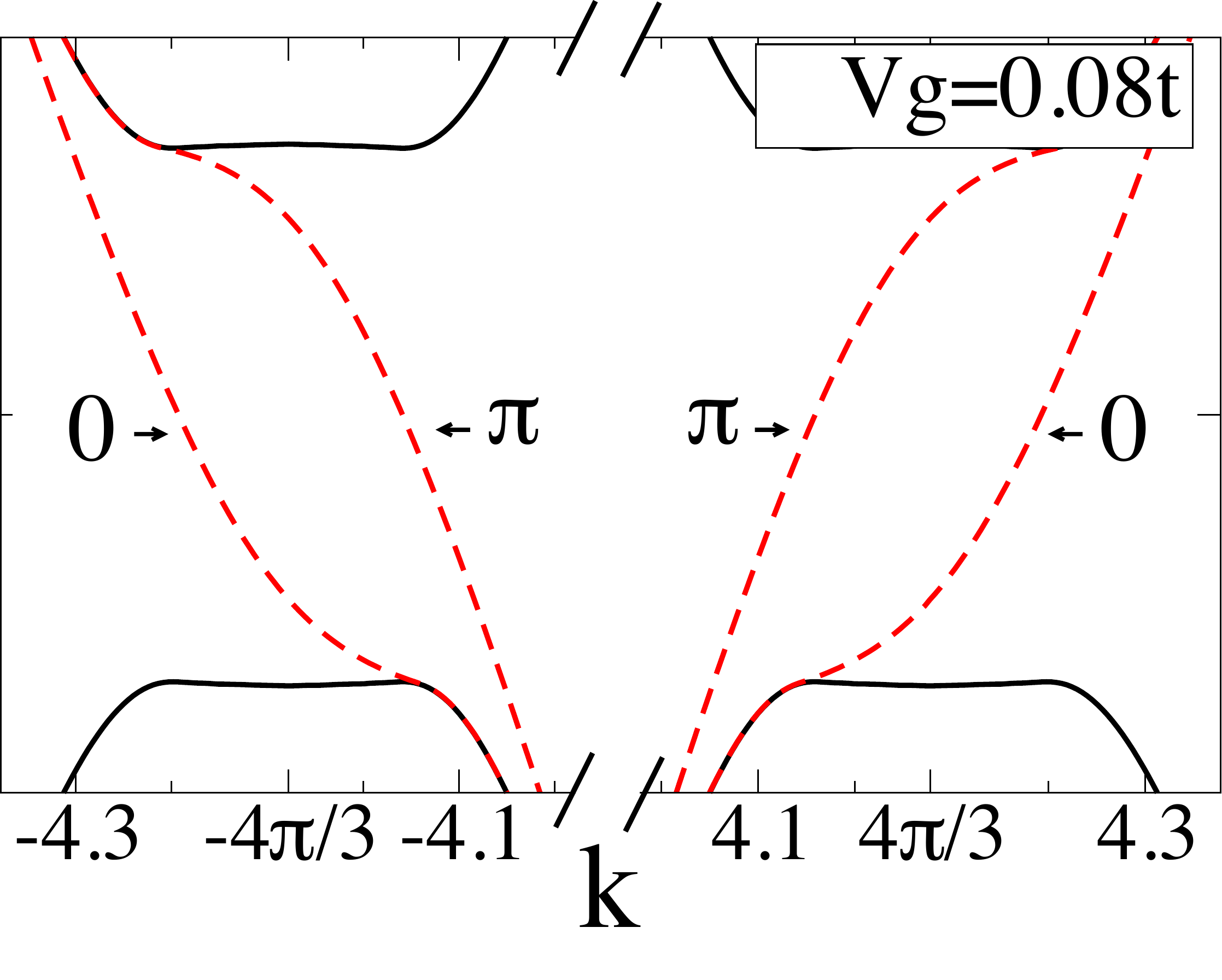} 
	\includegraphics[height=2.8cm]{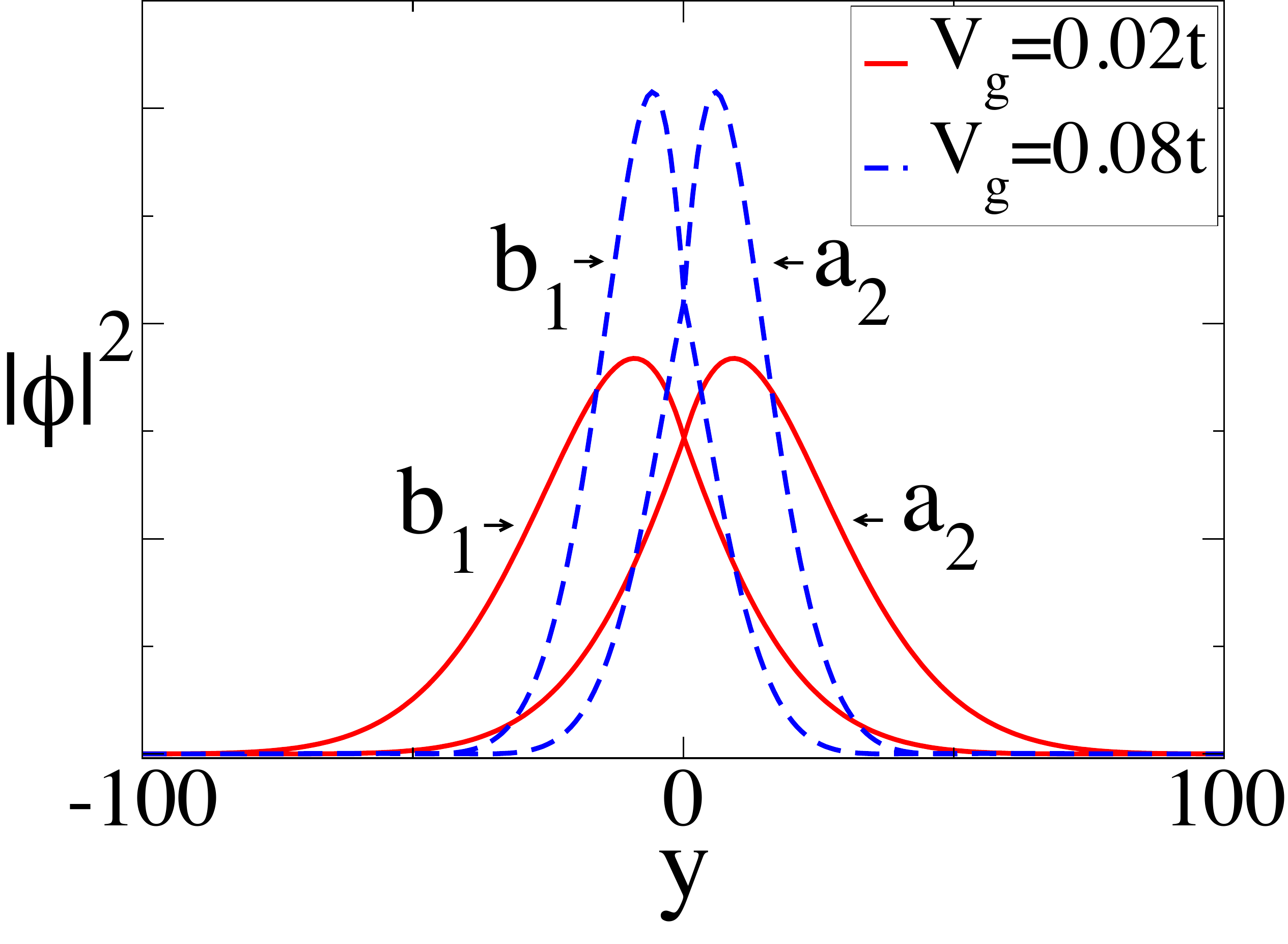} 
	\caption{Dispersion about the K-points with (a) $V_{g}=0.02t$ and (b) $V_{g}=0.08t$. Edge-mode bands are indicated by the labelled arrows and bulk-states by the hatched region. The modulus square of the zero-energy wavefunction of (c) the $0$-band at $V_g=0.02t$ and $V_g=0.08t$ (the $a_2$ and $b_1$ branches are exchanged for the $\pi$-band). }
\label{Fig:KinkDispersion}
\end{figure}

Solving the full tight binding model for a single kink yields the dispersion depicted in Fig.~\ref{Fig:KinkDispersion}. A single kink interface generates right moving subgap modes in one valley and two left moving modes in the opposite valley (labeled in red).  For an anti-kink profile, the dispersion is identical except the modes velocities are reversed in each valley.

In terms of the low energy, the eigenfunctions are then of the form \begin{eqnarray} \left( 
\begin{array}{cc}
	f(y) \\
	g(y) 
\end{array}
\right)_{0/\pi}=\left( 
\begin{array}{cc}
	f(y) \\
	f(-y) 
\end{array}
\right)_0,\left( 
\begin{array}{cc}
	f(y) \\
	-f(-y) 
\end{array}
\right)_\pi, 
\end{eqnarray}
 with corresponding eigenvalues of $+1$ and $-1$ of the operator $\mathcal{P}$, respectively.  Solutions with eigenvalues $+1$ with even symmetry belong to the lower $0$-band while solution with odd symmetry belong to the upper $\pi$-band.

Increasing the interlayer bias strength results in two important effects that can be seen qualitatively in Fig.~\ref{Fig:KinkDispersion}: i) the Fermi velocity of the two bands is enhanced, and ii) the wavefunctions become more confined to the interface.  With respect to the overall width of the wavefunction, a simple scaling analysis suggests that the wavefunction width should go as $l\sim (m^* V_g)^{-1} \sim \left(\frac{t}{\sqrt{V_g t_{\perp}}}\right)a$.

The large width of the wavefunction transverse to the wire direction strongly suppresses the bare backscattering terms due to electron-electron interactions.
A similar effect also seen in wide carbon nanotubes,\cite{White:1998} and it leads to a dominance
of forward scattering processes, where a simple bosonization analysis predicts a spin-charge
separated gapless Tomonaga-Luttinger liquid.
We thus expect a relatively large energy window where interactions drive the 1D
modes even in these `kink' modes in bilayer graphene into such a Luttinger liquid.
Remarkably, such a bosonization analysis arrives at a novel two-band Luttinger liquid with 
tunable mode velocities and tunable Luttinger parameters.\cite{Killi:2010}

Naively, it would appear that the localized kink states may not be robust because they are not topologically protected.  However, Qiao et al,\cite{Qiao:2011} performed an extensive study into various potentially detrimental mechanisms.  Through numerical conductance calculations and examining the LDOS, they showed the low energy states are remarkably robust to both short and long range disorder, and even to abrupt changes in the interface direction.  Although quantized conductance is not unlikely, the mean free path could be as large as a hundred microns for relatively clean samples. The mechanism which quantitatively leads to a strong suppression of 
backscattering is again the large wavefunction spread.

It has also been observed that the kink-modes are also quite robust when subjected to a magnetic field.\cite{Zarenia:2011,Zarenia:2011a,Wu:2012a}  This is due to the strong magnetic field induced confinement of the wavefunctions.\cite{Zarenia:2011}  Interestingly, by coupling a pair of coupled 
kink and antikink modes and applying a magnetic field, the current in the kink flows in one direction, opposite to the direction of flow in the antikink.  Moreover, all the dispersing modes have the same valley index, making this a potential valley filter.\cite{Wu:2012a,Rycerz:2007}  Recently, it has been proposed that at $\nu=0$ electron interactions form a charge density pattern in the vicinity of a kink state, which provides a key signature of quantum Hall ferromagnetism.\cite{Huang:2012}

Before turning our discussion back over to superlattices, we close this section by emphasizing that similar localized kink states are also expected to form naturally in the presence of charge impurities and in strongly correlated phases where the $Z2$ layer symmetry is spontaneously broken.  In the case of the former, charge impurities close to the surface of the sample can generate a local electric field strong enough to induce an interlayer bias.  In uniformly biased BLG or where there are multiple charge impurities in close vicinity (on opposite layers or with opposite charge), this can cause the interlayer bias to reverse, generating kink states -- a point to be elaborated on in the conclusion.  In the latter case, any state with spontaneously broken layer symmetry will naturally form domain walls separating regions with opposite interlayer bias.\cite{Nandkishore:2012,MacDonald:2012}  Again, kinks states are expected to from percolation networks that permeate throughout the bulk. 
%%%%%%%%%%%%%%%%%%%%%%%%%%%%%%%%%%%%%%%%%%%%%%%%%%%%%%%%%%%%%%%
\subsubsection{Electric Field Superlattice: Effective Model}
\begin{figure}
	[b] \centering 
	\includegraphics[height=2.8cm]{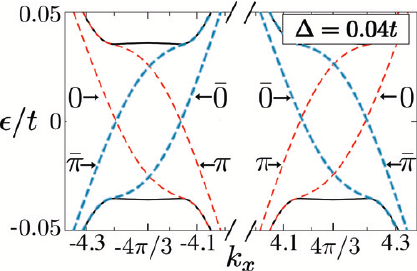} \quad \quad 
	\includegraphics[height=2.8cm]{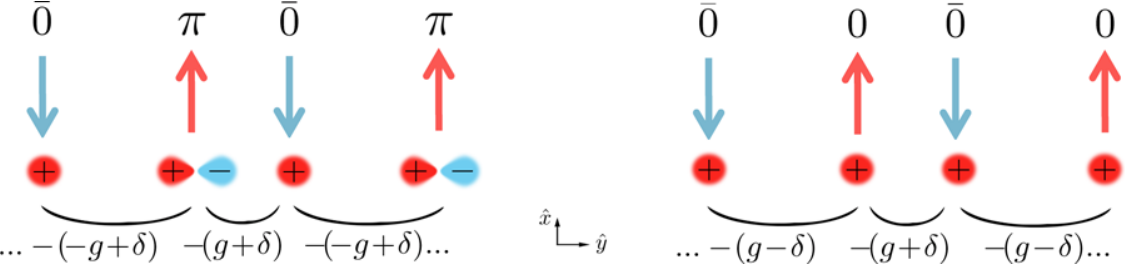} \caption{(color online) Left: Spectrum of isolated kink (thin, red) and anti-kink (thick, blue). Higher (lower) energy modes are labelled $\pi$ ($0$) at a kink and as $\bar{\pi}$ ($\bar{0}$) at an anti-kink. Right: Schematic of hopping between the $\pi-\bar{\pi}$ and $\bar{0}-\pi$ states.}  
	\label{Fig:WireDispersion}
\end{figure}

%%%%%%%%%%%%%%%%%%%%%%%%%%%%%%%%%%%%%%%%%%%%%%%%%%%%%%%%%%%%%%% \label{Sect:Model}
Equipped with an understanding of the properties of the soliton kink/anti-kink modes, it is now possible to describe how to construct a low energy effective model.~\cite{Killi:2011a} 
To begin, first consider the dilute limit where the period length is much longer than the characteristic spread of the soliton modes, (i.e.\ $\lambda>>l$).  Each kink supports two (ignoring spin) unidirectional dispersing soliton modes while each anti-kink supports two oppositely moving modes per valley, as shown in Fig.~\ref{Fig:WireDispersion}  The counterpropagation of the kink and anti-kink modes results in four band crossing points about each {\bf K}-point, two at zero energy between bands with the same symmetry ($0$-$\bar{0}$ and $\pi$-$\bar{\pi}$) and two at finite energy between modes with opposite symmetry ($\pi$-$\bar{0}$ and $0$-$\bar{\pi}$).  As described below, when the wavefunctions of neighbouring soliton modes couple (i.e.\ $\lambda~l$), Dirac cones precipitate precisely at the band crossing points.  

At energies and momenta in the the vicinity of any one of the band crossing points the system looks as if it were an array of 1D chiral `wires' lying along the kinks and anti-kinks of the SL. Each wire supports modes that flow in opposite direction to its two neighbors.  Now, as the wavefunctions of these modes begin to overlap the electrons can hop between neighbouring wires.

With this in mind, let us consider the $\pi$-$\bar{0}$ modes at zero energy and at a momentum $p^*_x$ (away from ${\bf K}$). The hopping between neighboring wires along $\hat{y}$ is then between states which have opposite velocities (since it is between a kink and an anti-kink edge state) and it is between a p-wave like state (${\cal P}$-odd) and an s-wave like state (${\cal P}$-even) (see Sec.~\ref{sect:Kink}).

Careful attention must be made to get the correct form of transfer integral that describes the hopping between the wires.  The sign of the hopping can be deduced by taking the wave functions of the $0$-band and $\pi$-band as having s-wave orbital and p-wave orbital character, respectively, and noting the sign of the overlap between the wires.  An illustrative picture for two of the band crossing points is shown on the right in Fig.~\ref{Fig:KinkDispersion}. Hence, the hopping between wires at a zero energy band crossing point is staggered and uniform for the finite energy band crossing points.  Further details are provided in our previous work.~\cite{Killi:2011a}

Using the index $n$ to label the wires, the interchain hopping parameter will then alternate as $(-1)^n g$ for equally spaced wires and as $g+\delta, -g+\delta$ (with $\delta < g$) if pairs of wires are closer to each other. Linearizing the dispersion at the crossing point, and letting $v_0$ denote the velocity of the linearized modes, 
\bea H(\px)=&v_0&\sum_n \left((-1)^n (p_x-p^*_x) c^{\dg}_{\px n}c_{\px n}\right) \nonumber \\
&-&\sum_n (g (-1)^n + \delta) \left(c^{\dg}_{\px n}c_{\px n+1} + h.c.\right) \eea 
where $p^*_x$ is the location of the $\pi-{\bar 0}$ crossing point in the single kink or antikink problem, and $c_{\px n}$ annihilates an electron on wire $n$ with momentum $\px$. Let $\xi(p_x)\equiv v_0 (p - p^*_x)$. Fourier transforming, we find $H(\px)=\sum'_\py \Psi^\dg(\py) {\bf \sigma} \cdot {\bf h}(\px) \Psi(\py)$, where ${\bf h}(\px)=\left(\xi(p_x) , -2 g \sin(\py), -2 \delta \cos(\py)\right)$, with $\Psi(\py) =(c_\py \, c_{\py+\pi})^T$, and $\sum'_\py$ runs over the MBZ. The dispersion is thus $E=\pm \sqrt{\xi^2(p_x) + 4 \delta^2 \cos^2(\py) + 4 g^2 \sin^2(\py)}$. Consequently, when $w=\lambda/2$, and the Hamiltonian commutes with ${\cal P}$, we have $\delta=0$ and a Dirac cone is generated at $(p^*_x, 0)$, consistent with numerical results. When $w \neq \lambda/2$, the Hamiltonian breaks ${\cal P}$ --- we then have $\delta \neq 0$, which leads to a gap $4\delta$. Similar arguments hold for the other zero energy band crossing points. The velocity of the Dirac fermions is highly anisotropic and depends on $g$, except along the SL direction where it inherits its value from the freestanding zero mode velocity --- this can be controlled by tuning the SL period and amplitude.

There are a number of particularly salient properties of the electric field SL when viewed from the `coupled wire' perspective. Although for different reasons than the chemical potential SL, its band structure is also expected to be quite resilient to disorder.  As discussed previously, the underlying soliton modes of the SL are exceptionally robust to disorder and fortifies the the band structure.  Moreover, since the individual 1D soliton modes are present over a wide range of interlayer biases, the Dirac spectrum of SL is persists for both weak and strong SL potentials.  In addition, the properties of the low energy Dirac fermions are very versatile.  Specifically, the velocity parallel and perpendicular to the modulation direction can be tuned \textit{independently} by first adjusting the SL strength and then the period length.  Furthermore, mass can be imparted to the fermions by breaking the $\cal{P}$-symmetry of the SL.  Interestingly, just as in polyacetylene, a domain wall between a gapped region with $w >\lambda/2$ and a gapped region with $w <\lambda/2$ leads to new subgap soliton modes. Since each kink/anti-kink is itself like a domain wall, these should be viewed as solitons in a soliton lattice.

%%%%%%%%%%%%%%%%%%%%%%%%%%%%%%%%%%%%%%%%%%%%%%%%%%%%%%%%%%%%%%%
%%%%%%%%%%%%%%%%%%%%%%%%%%%%%%%%%%%%%%%%%%%%%%%%%%%%%%%%%%%%%%%
\subsection{Magnetic Field Effects on 1D superlattices}\label{sect:Magnetic}
%%%%%%%%%%%%%%%%%%%%%%%%%%%%%%%%%%%%%%%%%%%%%%%%%%%%%%%%%%%%%%%
%%%%%%%%%%%%%%%%%%%%%%%%%%%%%%%%%%%%%%%%%%%%%%%%%%%%%%%%%%%%%%%
In this section, we review the effects of a magnetic field on the single particle properties of BLG subject to 1D SLs, but before doing so, it is useful to briefly review the LLs of intrinsic BLG.  The low energy model that describes BLG in the presence of a perpendicular magnetic field is obtained by replacing the momentum operator in Eqn.~\ref{Hred} with its canonical counterpart to take into account of the magnetic field effect.

The following results were obtained from effective two band model with the same gauge choice as before, $A=By\hat{x}$. The eigenvalues and the corresponding eigenvectors of the above Hamiltonian for the $s=+1$ valley in the absence of a SL are
\begin{eqnarray} &&\varepsilon_n={\rm sgn}(n)\sqrt{|n|(|n|-1)}\omega_c^2/t_{\perp},\nonumber\\
&&\phi_{n,k,+}(x,y)=\frac{e^{i k x}}{\sqrt{2 L}} \left( 
\begin{array}{c}
	\psi_{|n|,k}(y) \\
	- {\rm sgn}(n)\psi_{|n|-2,k}(y) 
\end{array}
\right), 
\label{Eq10}
\end{eqnarray}
with $|n| \geq 2$. In addition, there are two zero energy solutions that are feature the hallmark feature of the quadratic band touch point,\cite{McCann:2006,Novoselov:2006} \begin{eqnarray}
&&\varepsilon_1=0,\ \ \ \ \ \phi_{1,k,+} (x,y)=\frac{e^{ i k x}}{\sqrt{L}}\left( 
\begin{array}{c}
	\psi_{1,k} (y) \\
	0 
\end{array}
\right), \nonumber\\
&&\varepsilon_0=0,\ \ \ \ \ \phi_{0,k,+} (x,y)=\frac{e^{ ikx}}{\sqrt{L}}\left( 
\begin{array}{c}
	\psi_{0,k} (y) \\
	0 
\end{array}\right).
\end{eqnarray}
For $s=-1$, the corresponding eigenvectors are given by $\phi_{n,k,-} (x,y) = \sigma_x \phi_{n,k,+} (x,y)$. The full low energy LL wavefunctions thus take the form $\phi_{n,k,\pm} {\rm e}^{\pm i K_x x}$. and these serve as a good basis to study the magnetic field effect of bilayer graphene SLs.

%%%%%%%%%%%%%%%%%%%%%%%%%%%%%%%%%%%%%%%%%%%%%%%%%%%%%%%%%%%%%%%
\subsubsection{Chemical Potential Superlattice: Landau Levels and DC Conductivity}
\label{sect:ChemB}
\begin{figure}
	[th] \centering
	\includegraphics[width=0.6
	\textwidth]{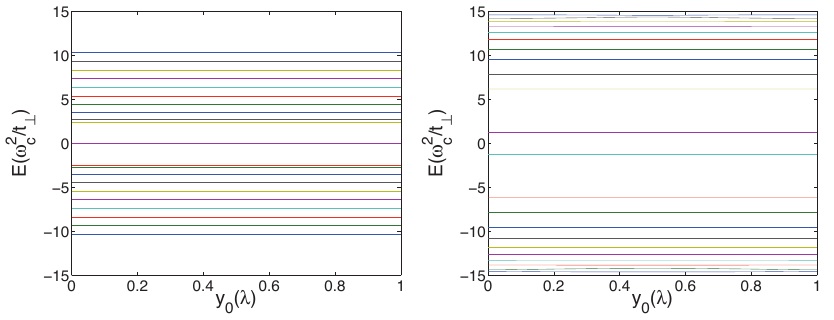}
	\includegraphics[width=.3
	\textwidth]{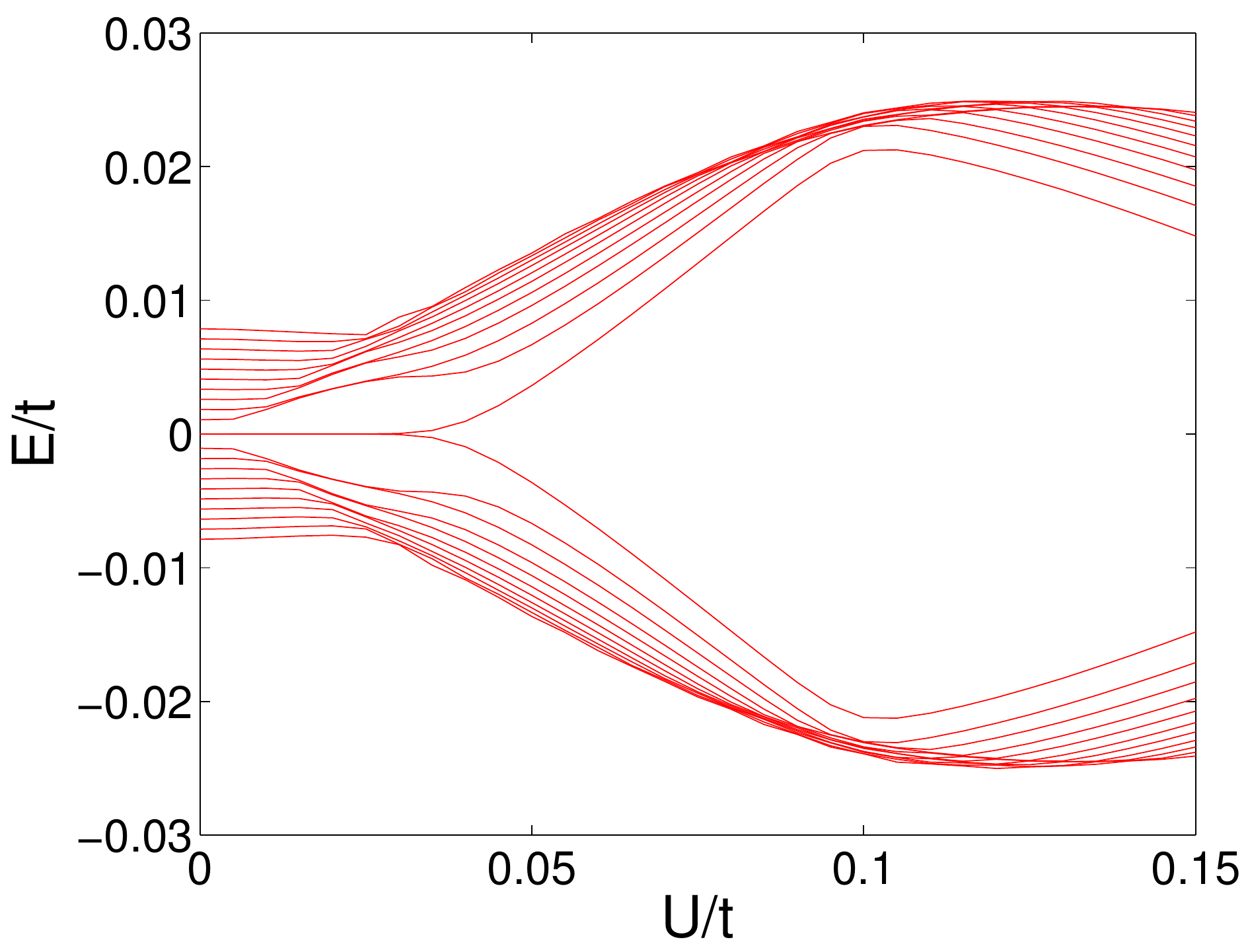}
	\caption{(Color online) Left and center panel: Energy spectrum of BLG subject to a chemical potential SL and a weak perpendicular magnetic field ($\ell_B=2\lambda$) for $V_0=0.01t$ and $V_0=0.04t$, respectively. Right panel: Evolution of low lying energy levels as a function of SL potential strength $U$, with $\ell_B=2\lambda$, $y_0=0$. In all cases, $\lambda=100a$, where $a=1.42$\AA. } \label{BLGSL_LL} 
\end{figure}

%%%%%%%%%%%%%%%%%%%%%%%%%%%%%%%%%%%%%%%%%%%%%%%%%%%%%%%%%%%%%%%
Just as for the Dirac cones derived from the SL in the single layer, evidence of Dirac fermion dispersion was shown to exist in the LL structure of 1D SL in BLG.\cite{Wu:2012a} The LL energy spectrum generated by a weak magnetic field is shown in the left panel of Fig.~\ref{BLGSL_LL} for $V_0=0.01t$. Degenerate pairs of energy levels that derive from the anistropic Dirac cones are present at zero energy ($n=0$) and at finite energy ($n=\pm 1$).  The higher LL, however, do come in degenerate pairs because it is only at lower energies the spectrum consists of two copies of Dirac cones.  The right panel shows that for $V_0=0.04t$, the zero energy LL levels are absent, consistent with the opening of a band gap.

The evolution of the LLs as the SL potential is increased, shown in Fig.~\ref{BLGSL_LL}, displays a definitive crossover from a non-relativistic to a relativistic regime, in addition to the opening of a bandgap.  As SL potential increases, the physics gradually becomes dominated by anisotropic Dirac cones, which can be seen from the appearance of doubly degenerate levels at nonzero energies.  The existence of a marked crossover can be qualitatively understood by considering the competition between the characteristic energy scales in these two regimes. In the absence of the SL, the low energy excitations are massive electrons with an effective mass $m^*=t_{\perp}/2v_F^2$ and have a cyclotron frequency, $\omega_c^{\prime}=eB/m^*c=1/m^*\ell_B^2$. On the other hand, the anisotropic Dirac points generated by SL have anisotropic Fermi velocities $v_y=\sqrt{2}\lambda|U({\bf Q})|/\pi$ and $v_x=2v_y$, where ${\bf Q}=\hat{y}2\pi/\lambda$,\cite{Killi:2011a} and have the characteristic energy scale of $\omega_c=\sqrt{2v_xv_y}/\ell_B$. Wu et al.\cite{Wu:2012a} estimated the crossover should occur around $U\sim 0.002t$, which is quite close to the value observed in 
Fig.~\ref{BLGSL_LL} (right figure).

Further increasing the SL potential, the doubly degenerate zero energy levels become gapped and all levels are pushed away from Dirac point. Surprisingly, at rather strong SL potential, $U\sim 0.22t$, zero energy LLs appear again, and all of the higher energy levels become doubly degenerate. This phenomenon can be understood from the result of Tan {\textit et al}.\cite{Tan:2011} As it has been shown, for a chemical potential SL, when SL potential is strong enough, anisotropic Dirac cones will show up again in the energy spectrum, which naturally leads to the zero energy LL in the presence of a magnetic field. As shown elsewhere,\cite{Tan:2011} there can be up to four Dirac points in the spectrum. For even stronger chemical potential SLs, the degeneracy of the zero energy LL reduces to two, consistent with two of the Dirac cones becoming gapped as discussed in Sec.~\ref{sect:BLG_electric}.
\begin{figure}
	[tb] 
	\centering
	\includegraphics[width=.78
	\textwidth]{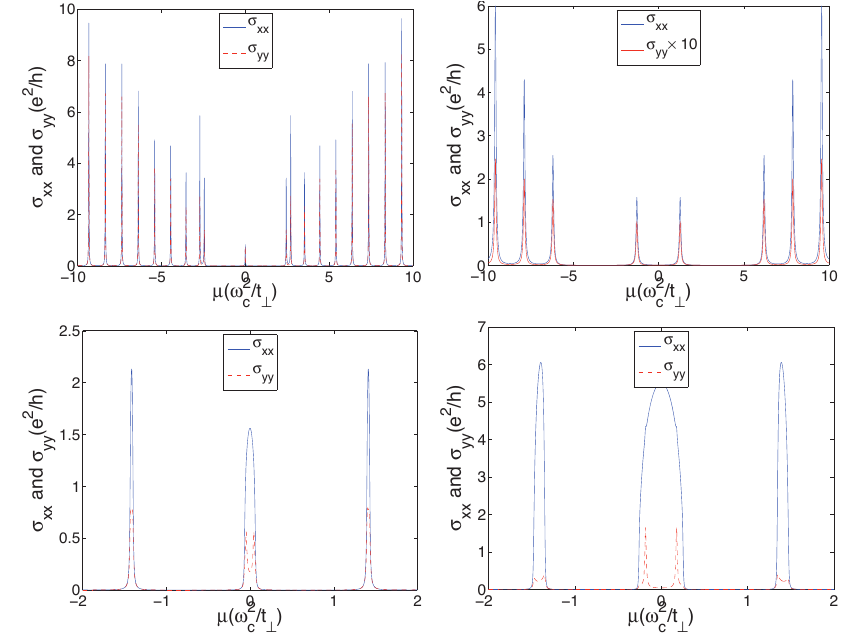}  
	\caption{(Color online) Diagonal dc conductivities for bilayer graphene for a chemical potential SL with different strengths $V_0$, and magnetic fields $B$. The conductivity is shown for weak field ($\ell_B=2\lambda$, top panels) and intermediate field ($\ell_B=0.2\lambda$, bottom panels). Left panels (a,b) correspond to $V_0=0.01t$, and right panels (c,d) correspond to $V_0=0.04t$. The conductivities show anisotropy, where $\sigma_{xx}$ is always larger than $\sigma_{yy}$, in contrast to anisotropy reversal in MLG SLs. } \label{BLGSL_diagonal} 
\end{figure}
Fig. \ref{BLGSL_diagonal} shows the dc diagonal conductivity of chemical potential SLs, where we again see anisotropy similar to the single layer case. However, in contrast to the monolayer, the direction with largest conductivity does not reverse when the magnetic field strength is tuned, and so the anisotropy cannot be tuned. For weak fields, the transport anisotropy is directly determined by the anisotropy of the Dirac cones. In Ref. \cite{Killi:2011a}, the extent anisotropy was determined by treating the SL potential as a perturbation.  It was calculated to be $v_x \simeq 2v_y$ for the emergent Dirac cones, consistent with the observation that the conductivity in the $\hat{x}$ direction, $\sigma_{xx}$, is larger than $\sigma_{yy}$ in a weak magnetic field.   As for intermediate magnetic field, due to the dispersion of the energy bands, the average velocity in the $\hat{x}$ direction is not zero, $\langle \hat{v}_x\rangle \neq 0$. On the other hand, $\langle \hat{v}_y\rangle$ is always equal to zero. This means that $\sigma_{xx}$ will acquire intra-LL contributions, while $\sigma_{yy}$ is mainly determined by inter-LL contributions and is thus small compared to $\sigma_{xx}$.

%%%%%%%%%%%%%%%%%%%%%%%%%%%%%%%%%%%%%%%%%%%%%%%%%%%%%%%%%%%%%%%
\subsubsection{Electrical Field Superlattice: Landau Levels and DC Conductivity}\label{sect:ElectricB}
%%%%%%%%%%%%%%%%%%%%%%%%%%%%%%%%%%%%%%%%%%%%%%%%%%%%%%%%%%%%%%%
\begin{figure}[bt]
	\centering 
	\includegraphics[height=3.1cm]{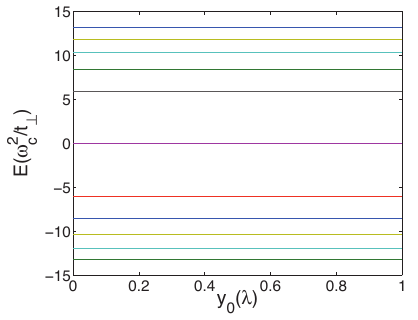} 
	\includegraphics[height=3cm]{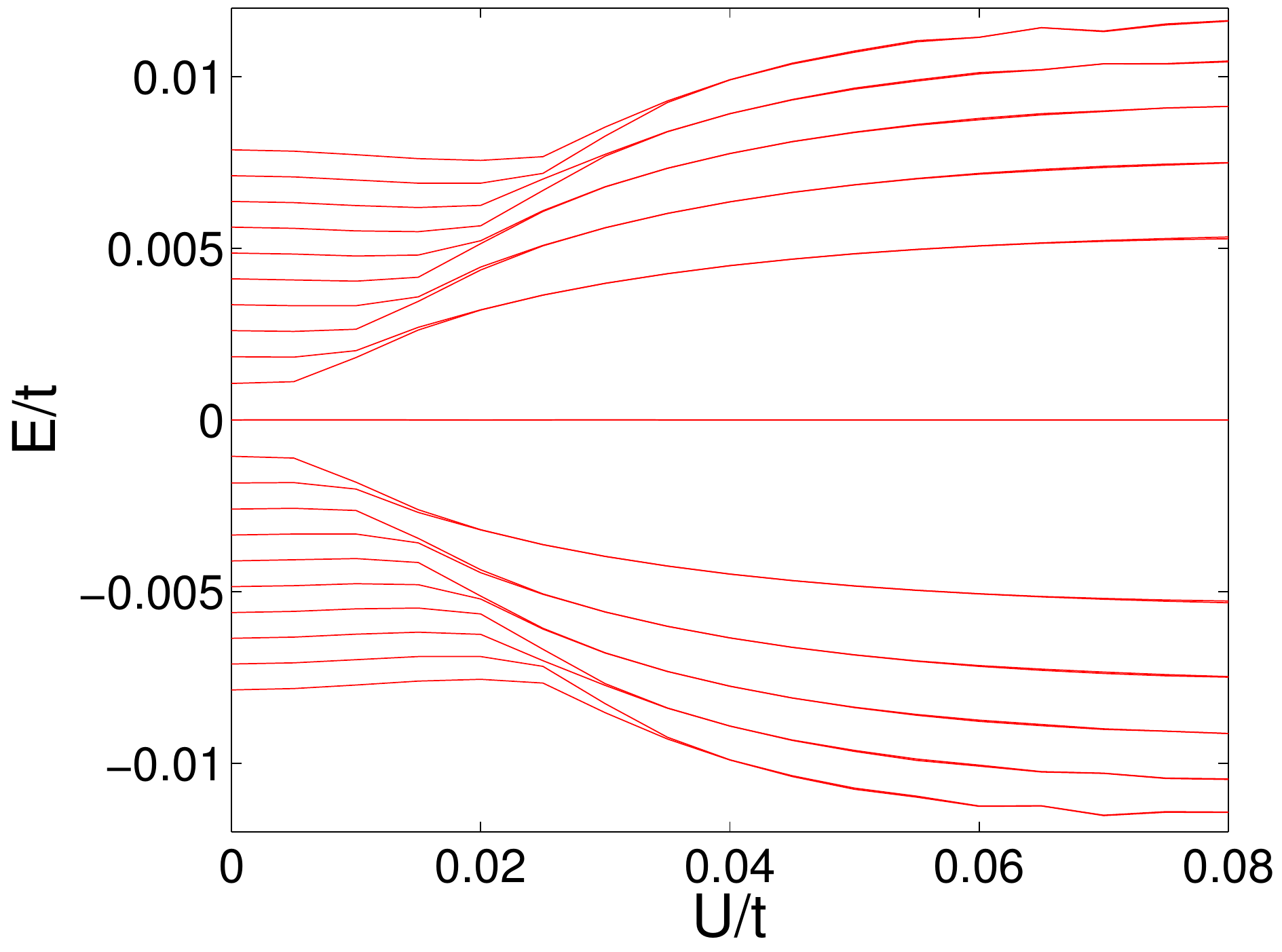}
	\includegraphics[height=3.1cm]{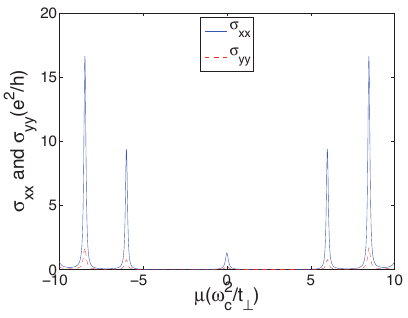} 
	\caption{Left: Energy spectrum of an electric field superlattice, in a weak magnetic field. Centre: Evolution of low lying energy levels in an electric field SL as a function of SL potential strength $U$ with $y_0=0$.   Right: DC conductivity.  In all plots, $\ell_B=2\lambda$, , $\lambda=100a$ where $a=1.42$\AA.} \label{delta} 
\end{figure}

In a 1D symmetric electric field SL, there are always two zero energy Dirac points present in the spectrum, which results from the coupling of 1D zero modes at kink/antikink of the SL potential profile. Wu et al.\cite{Wu:2012a} argued that this implies that when a weak magnetic field is applied, doubly degenerate zero energy levels should appear at the Dirac point. Indeed, as can be verified from the LLs shown in the left panel and the evolution of the LLs in the centre panel of Fig. \ref{delta}, these two levels are always present at zero energy and are independent of the SL potential strength.

The authors also showed that for a SL potential chosen to be $V_0=0.03t$, nearly degenerate levels even appear at nonzero energies (Fig. \ref{delta}). These correspond to the LLs derived from anisotropic Dirac points, up to $n=\pm 4$. From the left panel of Fig. \ref{delta}, it is more clear that at strong SL potential, physics is strongly dominated by the Dirac points, where higher energy levels become doubly degenerate and resemble the higher LLs of the Dirac cones. When SL potential is weak, equally spaced LLs are recovered, as in the chemical potential SL, which also indicates a nonrelativistic to relativistic crossover at certain SL strength. Remarkably, and different from the chemical potential SL case, the relativistic behavior survives to higher energies as SL potential increases, which means the linear approximation description of Dirac cones works in a larger energy range. This is consistent with earlier result.\cite{Killi:2010,Killi:2011a} Therefore, as the SL potential increases, the energy range where the Dirac cone approximation is valid also increases, which leads to the robust relativistic physics at large SL potential.  In addition, evidence for the large anisotropy of Dirac cones discussed previously can be readily seen in the DC conductivity shown in the right panel of Fig.~\ref{delta}.

\section{Concluding remarks}
We have explored, here, the rich physics associated with slowly modulated potentials in monolayer
and bilayer graphene. The resulting dispersions and magnetotransport properties of such
superlattices can be explored experimentally by engineering gates
to pattern suitable superlattice potentials. Indeed, such periodic modulated potentials for
Dirac fermions have also been explored in recent scanning tunneling spectroscopy studies
the surface states of Bi$_2$Te$_3$, a topological insulator,
where the modulation arises from the periodic
buckling of the crystal structure.\cite{Madhavan:2012}
Going beyond such slowly varying potentials, there
has been a lot of recent interest in graphene on hexagonal Boron Nitride substrates, where
lattice mismatch leads to Moir\'e patterns - such potentials imparted by the hBN substrate
have slow as well as fast sublattice scale components, leading to new physics which is still
being explored.\cite{Sachs:2011,Ortix:2011,Jung:2012,Kindermann:2012a} Experimentally, tunneling studies indicate the emergence of new finite energy
Dirac points in this case,\cite{Yankowitz:2012} and further studies in this area would be valuable. Turning to a 
different aspect of such potential modulations, we note that the issue of transport in bilayer
graphene under a uniform bias is likely to be impacted by the presence of charged impurities
in the substrate.\cite{Taychatanapat:2010,Rossi:2011,Rutter:2011,Abergel:2011,Abergel:2012} The electric field of such an impurity could locally reverse the applied bias,
leading to a ring around the impurity site which can trap midgap states.\cite{Xavier:2010} While the ring size is
likely to be small, on the nanometer scale, the trapped states would have a large wavefunction
spread due to the very small effective mass associated with the quadratic band touching point
in bilayer graphene. In this case, we expect the low temperature transport could be via
hopping conduction between such ``ring'' localized states, while the observed small activation
in biased samples\cite{Taychatanapat:2010,Yan:2010} could result from such hopping conduction getting gapped out due to the
charging energy of such ``ring'' states. Rough estimates show that the wavefunction spread
could be on the order of $50$nm, while the charging energy could be $\sim 10$meV,
comparable to gaps observed in transport measurements.\cite{Taychatanapat:2010,Yan:2010}  
The study of a network model of 
such ``ring'' states is likely to yield new insights into transport mechanisms in bilayer graphene.
In summary, the study of periodic and random potential modulations in graphene and bilayer
graphene is a rich and growing field with many open questions and we invite the reader to 
join us on this exciting journey.

\section{Acknowledgments} We thank Jin-Luo Cheng and Jeil Jung for discussions. This 
research was supported by NSERC of Canada, an Ontario Early Researcher Award, and
a University of Waterloo start-up grant (SW). We also acknowledge the hospitality of the
International Center for Theoretical Sciences, Bangalore, where some of this research was
completed.

\end{document}